\documentclass[a4paper,fleqn,usenatbib]{mnras}

\usepackage{newtxtext,newtxmath}

\usepackage[T1]{fontenc}
\usepackage{ae,aecompl}

\usepackage{graphicx}	
\usepackage{amsmath}	
\usepackage{gensymb}

\defcitealias{NE2001}{NE2001}
\defcitealias{YMW16}{YMW16}
\defcitealias{4FGL}{4FGL}

\title[Discovery of PSR J2039$-$5617]{Einstein@Home Discovery of the Gamma-ray Millisecond Pulsar PSR J2039$-$5617 Confirms Its Predicted Redback Nature}

\author[C. J. Clark et al.]
       {
         C.~J.~Clark$^{1}$\thanks{E-mail: colin.clark-2@manchester.ac.uk},
         L.~Nieder$^{2,3}$\thanks{E-mail: lars.nieder@aei.mpg.de},
         G.~Voisin$^{1,4}$,
         B.~Allen$^{2,5,3}$,
         C.~Aulbert$^{2,3}$,
         O.~Behnke$^{2,3}$, \newauthor 
         R.~P.~Breton$^{1}$, 
         C.~Choquet$^{6}$, 
         A.~Corongiu$^{7}$, 
         V.~S.~Dhillon$^{8,9}$, 
         H.~B.~Eggenstein$^{2,3}$, \newauthor
         H.~Fehrmann$^{2,3}$, 
         L.~Guillemot$^{10,11}$, 
         A.~K.~Harding$^{12}$, 
         M.~R.~Kennedy$^{1}$, \newauthor
         B.~Machenschalk$^{2,3}$, 
         T.~R.~Marsh$^{13}$,
         D.~Mata~S\'anchez$^{1}$, 
         R.~P.~Mignani$^{14,15}$, \newauthor
         J.~Stringer$^{1}$, 
         Z.~Wadiasingh$^{12}$, 
         J.~Wu$^{16}$
         \medskip\\
$^1$ Jodrell Bank Centre for Astrophysics, Department of Physics and Astronomy, The University of Manchester, M13 9PL, UK\\
$^2$ Albert-Einstein-Institut, Max-Planck-Institut f\"ur Gravitationsphysik, D-30167 Hannover, Germany\\
$^3$ Leibniz Universit\"at Hannover, D-30167 Hannover, Germany\\
$^4$ LUTH, Observatoire de Paris, PSL Research University, 5 Place Jules Janssen, 92195 Meudon, France\\
$^5$ Department of Physics, University of Wisconsin-Milwaukee, P.O. Box 413, Milwaukee, WI 53201, USA\\
$^6$ R\'{e}sidence Le Dauphin\'{e}, rue Jean Bleuzen, Vanves, France\\
$^7$ INAF - Osservatorio Astronomico di Cagliari, Via della Scienza 5, I-09047 Selargius (CA), Italy\\
$^8$ Instituto de Astrof\'isica de Canarias (IAC), E-32800, La Laguna, Tenerife, Spain\\
$^9$ Department of Physics and Astronomy, University of Sheffield, Sheffield S3 7RH, UK\\
$^{10}$ Laboratoire de Physique et Chimie de l'Environnement et de l'Espace -- Universit\'e d'Orl\'eans / CNRS, F-45071 Orl\'eans Cedex 02, France\\
$^{11}$ Station de radioastronomie de Nan\c{c}ay, Observatoire de Paris, CNRS/INSU, F-18330 Nan\c{c}ay, France\\
$^{12}$ NASA Goddard Space Flight Center, Greenbelt, MD 20771, USA\\
$^{13}$ Department of Physics, University of Warwick, Coventry CV4 7AL, UK\\
$^{14}$ INAF-Istituto di Astrofisica Spaziale e Fisica Cosmica Milano, via E. Bassini 15, I-20133 Milano, Italy\\
$^{15}$ Janusz Gil Institute of Astronomy, University of Zielona G\'ora, ul Szafrana 2, 65-265, Zielona G\'ora, Poland \\
$^{16}$ Max-Planck-Institut f\"ur Radioastronomie, Auf dem H\"ugel 69, D-53121 Bonn, Germany
       }
\date{Accepted 2020 September 27. Received 2020 September 25; in original form 2020 June 28}

\pubyear{2020}

\begin{document}
\label{firstpage}
\pagerange{\pageref{firstpage}--\pageref{lastpage}}
\maketitle

\begin{abstract}
The \textit{Fermi} Large Area Telescope gamma-ray source 3FGL J2039.6$-$5618
contains a periodic optical and X-ray source that was predicted to be a
``redback'' millisecond pulsar (MSP) binary system. However, the conclusive
identification required the detection of pulsations from the putative MSP. To
better constrain the orbital parameters for a directed search for gamma-ray
pulsations, we obtained new optical light curves in 2017 and 2018, which
revealed long-term variability from the companion star. The resulting orbital
parameter constraints were used to perform a targeted gamma-ray pulsation search
using the \textit{Einstein@Home} distributed volunteer computing system. This
search discovered pulsations with a period of $2.65$\,ms, confirming the source
as a binary MSP now known as PSR~J2039$-$5617.
    Optical light curve modelling is
complicated, and likely biased, by asymmetric
  heating on the companion star and long-term variability, but we find an inclination $i\gtrsim
$ 60\,$\degr$, for a low pulsar mass between $1.1\,M_{\odot} <
M_{\rm psr} < $ 1.6 $M_{\odot}$, and a companion
  mass of $0.15$--$0.22\,M_{\odot}$, confirming the redback
  classification. Timing the gamma-ray pulsations also revealed significant
variability in the orbital period, which we find to be consistent with
quadrupole moment variations in the companion star, suggestive of convective
activity. We also find that the pulsed flux is modulated at the orbital period,
potentially due to inverse Compton scattering between high-energy leptons in the
pulsar wind and the companion star's optical photon field.
\end{abstract}


\section{Introduction}
Millisecond pulsars (MSPs) are old neutron stars that have been spun-up to
millisecond rotation periods by the accretion of matter from an orbiting
companion star \citep{Alpar1982+Recycling}. The most compelling evidence for
this ``recycling'' scenario comes from the discovery of three transitional MSPs,
which have been seen to switch between rotationally powered MSP and
accretion-powered low-mass X-ray binary (LMXB) states
\citep{Archibald2009+J1023,Papitto2013+M28I,Bassa2014+J1227,Stappers2014+J1023}. In
their rotationally powered states, these transitional systems all belong to a
class of interacting binary MSPs known as ``redbacks'', which are systems
containing an MSP in orbit with a low-mass ($0.1\,M_{\odot} \lesssim M
\lesssim 0.4\,M_{\odot}$) non-degenerate companion star
\citep{Roberts2013+Spiders}. Redbacks, and the closely related ``black widows''
(which have partially degenerate companions with $M \lesssim
0.05\,M_{\odot}$), are named after species of spiders in which the
heavy females have been observed to consume the smaller males after mating,
reflecting the fact that the lighter companion stars are being destroyed by the
pulsar's particle wind and/or intense high-energy radiation.

Until recently, only a handful of these ``spider'' systems had been found in
radio pulsar surveys of the Galactic field. This is most likely due to the
ablation phenomenon which gives redbacks and black widows their nicknames:
plasma from the companion can eclipse, scatter and disperse the MSP's radio
pulsations for large fractions of an orbit
\citep[e.g.,][]{Ray2013+J1311,Deneva2016+J1048}, causing these pulsars to be
easily missed in radio pulsar surveys. In addition, traditional ``acceleration''
search methods for binary pulsars \citep{Ransom2002+Fourier} are only optimal
when the integration time is $\lesssim 10\%$ of the orbital period, leading to
an additional sensitivity loss to spiders, which often have orbital periods
of just a few hours.

Fortunately, gamma-ray emission from an MSP does not suffer from strong propagation effects from intrabinary plasma structures. A new route for binary MSP discoveries
therefore appeared with the launch of the \textit{Fermi Gamma-ray Space Telescope} in
2008. The on-board Large Area Telescope (LAT) discovered gamma-ray pulsations
from a number of known MSPs shortly after launch
\citep{Abdo2009+LATMSPs}. Targeted radio observations of unidentified, but
pulsar-like \textit{Fermi}-LAT sources have since discovered more than 90 new
MSPs, more than a quarter of all known MSPs in the Galactic
field\footnote{\url{http://astro.phys.wvu.edu/GalacticMSPs/GalacticMSPs.txt}}. A
disproportionately large fraction of these are spiders that had been missed by
previous radio surveys \citep{Ray2012+PSC}.

In addition to the large number of radio-detected spiders found in
\textit{Fermi}-LAT sources, a growing number of candidate spiders have been
discovered through searches for optical and X-ray counterparts to gamma-ray
sources
\citep[e.g.,][]{Romani2014+J1653,Strader2014+J0523,Halpern2017+J0838,Salvetti2017+PSRCands,Li2018+J0954}.
In a few cases, the MSP nature of these sources was confirmed by the
detection of radio or gamma-ray pulsations \citep{Pletsch2012+J1311,Ray2020+J2339}, however most of these candidates remain
unconfirmed.

To overcome the difficulties in detecting spider MSPs in radio pulsation
searches, it is possible to directly search for gamma-ray pulsations in the LAT
data. In contrast to searches for isolated MSPs, which can be detected in
gamma-ray pulsation searches without any prior knowledge of the parameters
\citep{Clark2018+EAHMSPs}, gamma-ray pulsation searches for binary MSPs require
tight constraints on the orbital parameters of the candidate binary system to
account for the orbital Doppler shift \citep{Pletsch2012+J1311}, which would
smear out the pulsed signal if not corrected for. This in turn requires
long-term monitoring of the companion star's optical light curve to measure the
orbital period with sufficient precision, and spectroscopic radial velocity
measurements and/or light curve modelling to tie the photometric light curve to
the pulsar's kinematic orbital phase. Prior to this work, such searches have
been successful only twice \citep{Pletsch2012+J1311,Nieder2020+J1653}, with both
MSPs being extremely compact black widow systems with small orbital Doppler
modulations.

\citet{Salvetti2015+J2039} and \citet{Romani2015+J2039} discovered a
high-confidence candidate redback system in the bright, pulsar-like gamma-ray
source 3FGL~J2039.6$-$5618 \citep{3FGL}. This source is now known as
4FGL~J2039.5$-$5617 in the latest \textit{Fermi}-LAT Fourth Source Catalog
\citep[hereafter \citetalias{4FGL},][]{4FGL}. This system (which we refer to
hereafter as J2039) contains a periodic X-ray and optical source with orbital
period $P_{\rm orb} \approx 5.5$\,hr. The optical light curve exhibits two
``ellipsoidal'' peaks, interpreted as a tidally distorted companion star in an intense gravitational field being viewed from the side, where its projected surface area is highest. These peaks have unequal amplitudes, indicating a
temperature difference between the leading and trailing sides of the
star. Despite the high likelihood of this source being a redback system, the
pulsar remained undetected in repeated observations attempting to detect its
radio pulsations by \citet{Camilo2015+Parkes}.

On 2017 June 18, we took new observations of J2039 with the ULTRACAM
\citep{Dhillon2007+ULTRACAM} high-speed multi-band imager on the 3.5m New
Technology Telescope (NTT) at ESO La Silla. The goal of these observations was
to refine the orbital period uncertainty by phase-aligning a new orbital light curve with the
2014 GROND observations from \citet{Salvetti2015+J2039}. However, we found that
the optical light curve had changed significantly. Further observations obtained on 2018 June 02 also found a light
curve that differed from the first two. This variability, similar to that
discovered recently in other redback pulsars
\citep{vanStaden2016+SpottyRB,Cho2018+VariableRBs}, poses challenges for
obtaining reliable estimates of the physical properties such as the binary
inclination angle and pulsar mass via optical light curve modelling
\citep[e.g.,][]{Breton2012+Icarus,Draghis2019+BWs}.

Using constraints on the pulsar's orbital period and epoch of ascending node from preliminary models fit to the optical data, we performed a
gamma-ray pulsation search using the \textit{Einstein@Home} distributed
volunteer computing system \citep{Knispel2010+EAH,Allen2013+EAH}, which finally
identified the millisecond pulsar, now named PSR~J2039$-$5617, at the heart of
the system.

In this paper, we present the detection and timing of gamma-ray pulsations from
PSR~J2039$-$5617, and our new optical observations of the system. The paper is
organised as follows: in Section \ref{s:archival} we review the literature on
recent observations of the system to update our knowledge of its properties;
Section \ref{s:fermi} presents updated analysis of \textit{Fermi}-LAT gamma-ray
observations of 4FGL~J2039.5$-$5617, and describes the gamma-ray pulsation
search, discovery and timing of PSR~J2039$-$5617; in Section \ref{s:optical} we
describe the newly obtained optical data, and model the optical light curves to
estimate physical properties of the system and investigate the observed
variability; in Section \ref{s:discussion} we discuss the newly clarified
picture of PSR~J2039$-$5617 in the context of recent observations of redback
systems; and finally a brief summary of our results is given in Section
\ref{s:conclusions}.

Shortly after the discovery of gamma-ray pulsations reported in this paper, the
initial timing ephemeris was used to fold existing radio observations taken by
the CSIRO Parkes radio telescope. The resulting detections of radio pulsations
and orbital eclipses are presented in a companion paper (Corongiu, A. et
al. 2020, MNRAS, accepted.), hereafter Paper II.

\section{Summary of previous literature}
\label{s:archival}
The periodic optical counterpart to 4FGL~J2039.5$-$5617 was discovered by
\citet{Salvetti2015+J2039} and \citet{Romani2015+J2039} in photometric
observations of the gamma-ray source region taken over three nights on 2014 June
16--18 with GROND \citep{Greiner2008+GROND} on the ESO/MPG 2.2m telescope on La
Silla. These observations covered SDSS $g^\prime, r^\prime, i^\prime$, and
$z^\prime$ optical filters in simultaneous $115$\,s exposures, and $H$, $J$, and
$K$ near infrared filters in simultaneous $10$\,s exposures. For consistency
with the new optical light curves presented in this paper, we re-reduced the
optical observations but chose not to include
the infrared observations, which were not compatible with our reduction
pipeline. These observations revealed a double-peaked light curve typical of
redback systems, but with the peak corresponding to the companion's ascending
node brighter and bluer than that of the descending node. This requires the
trailing side of the star to be hotter than the leading side, perhaps due to
heating flux being redirected by an asymmetric intra-binary shock
\citep[e.g.,][]{Romani2016+IBS}, or due to the presence of cold spots on the
leading edge \citep[e.g.,][]{vanStaden2016+SpottyRB}.

\citet{Salvetti2015+J2039} and \citet{Romani2015+J2039} also analyzed X-ray
observations of J2039 taken by \textit{XMM-Newton}. These data had insufficient
time resolution to test for millisecond X-ray pulsations, but did reveal a
periodic ($\sim5.5$\,hr) modulation in the X-ray flux, which the authors
identified as likely being due to synchrotron emission from particles
accelerated along an intra-binary shock, commonly seen in black widow and
redback systems. However, without long-term timing to precisely measure the
orbital period the authors were unable to unambiguously phase-align the optical
and X-ray light curves.
The Catalina Surveys Southern Periodic Variable
Star Catalogue \citep{Drake2017+CSSPVC} includes 223 photometric observations of
J2039 between 2005 and 2010. While the uncertainties on these unfiltered data are
too large for a detailed study of the light curve over these 5 years, the
underlying periodicity is clearly recovered by a 2-harmonic Lomb Scargle
periodogram, which reveals a significant signal with an orbital period of
$P_{\rm orb} = 0.227980(1)$\,d with no significant aliases.
Folding at this period shows that the X-ray modulation peaks at the putative
pulsar's inferior conjunction, indicating that the shock wraps around the
pulsar. This scenario requires the companion's outflowing wind to overpower the
pulsar wind \citep[see e.g.][]{Romani2016+IBS,Wadiasingh2017+Shocks}.

Using 9.5\,yr of \textit{Fermi}-LAT data, \citet{Ng2018+J2039_orb} discovered
that the gamma-ray emission from J2039 contains a component below $3$\,GeV that
is modulated at the orbital period, peaking around the companion star's inferior
conjunction, i.e. half an orbit out of phase with the X-ray modulation. This
phase offset rules out synchrotron emission from particles accelerated along the
shock front as an origin for the gamma-ray flux, as such a component would occur
at the same orbital phase as the X-ray modulation. Instead,
\citet{Ng2018+J2039_orb} propose that this component is produced by inverse
Compton scattering between the pulsar's high-energy particle wind and the
companion star's optical photon flux. Such a component would be strongest if our
line of sight to the pulsar passes close to the limb of the companion star,
suggesting an intermediate inclination angle $i \sim 80 \degr$.

\citet{Strader2019+RBSpec} obtained spectroscopic observations with the Goodman
Spectrograph \citep{Clemens2004+Goodman} on the SOAR telescope. The spectra
suggest a mid-G-type companion star, with temperature $T \approx 5500$\,K and
variations of up to $\pm200$\,K across the orbit attributed to heating from the pulsar. The spectroscopy also revealed
a single-line radial velocity curve whose semi-amplitude of $K_{\rm c} =
324\pm5$\,km\,s$^{-1}$ implies an unseen primary with a minimum mass $M > 0.8
M_{\odot}$. 
\citet{Strader2019+RBSpec} modelled the GROND light curve, incorporating two
large cold spots on the outer face of the companion star to account for the
light curve asymmetry, and found an inclination angle $i \sim 55 \degree$, from
which they deduce a heavy neutron star primary with $M \gtrsim 1.8 M_{\odot}$.

The optical counterpart is also covered in the Second \textit{Gaia} Data Release
\citep[DR2,][]{GAIA2016,GAIA2018+DR2}.
Using Equation (2) of \citet{GAIA2010+Phot}, the \textit{Gaia} DR2 colour
$G_{\rm BP} - G_{\rm RP} = 1.02$ implies an effective temperature of $T_{\rm
  eff} = 5423 \pm 249$\,K, consistent with the spectroscopic temperature
measured by \citet{Strader2019+RBSpec}. The \textit{Gaia} DR2 also provides a
marginal parallax detection ($\varpi = 0.40 \pm 0.23$\,mas) for a minimum ($95\%$ confidence) distance of $d > 1.2$\,kpc, and a total proper motion of $\mu = 15.51 \pm 0.26$\,
mas~yr$^{-1}$, corresponding to a distance-dependent transverse velocity of $v(d) \approx 75\,
(d/1\,\textrm{kpc})$\,km~s$^{-1}$. The systemic velocity (the radial velocity of the binary centre of mass) measured from optical spectroscopy by \citet{Strader2019+RBSpec} is just $6\pm3\,$km~s$^{-1}$ indicating that the 3D velocity vector is almost entirely transverse.

\section{Gamma-ray Observations}
\label{s:fermi}
To update the gamma-ray analysis of J2039 from previous works
\citep{Salvetti2015+J2039,Ng2018+J2039_orb}, we selected \texttt{SOURCE}-class
gamma-ray photons detected by the \textit{Fermi} LAT between 2008 August 04 and
2019 September 12. Photons were included from within a $15\degree$ region of
interest (RoI) around J2039, with energies greater than $100\,$MeV, and with a
maximum zenith angle of $90\degree$, according to the ``Pass 8''
\texttt{P8R3\_SOURCE\_V2} \citep{Pass8,Bruel2018+P305} instrument response
functions (IRFs) \footnote{See \url{https://fermi.gsfc.nasa.gov/ssc/data/analysis/LAT_essentials.html}}.

We first investigated the gamma-ray spectral properties of
4FGL~J2039.5$-$5617. We used the \citetalias{4FGL} catalogue as an initial model
for the RoI, and used the \texttt{gll\_iem\_v07.fits} and
\texttt{iso\_P8R3\_SOURCE\_V2\_v1.txt} models to describe the Galactic and
isotropic diffuse emission, respectively. We replaced 4FGL~J2039.5$-$5617 in the
RoI model with a point source at the \textit{Gaia} DR2 position of the optical
source. To model the source spectrum, we used a subexponentially-cutoff
power-law spectrum typical for gamma-ray pulsars \citepalias{4FGL},
\begin{equation}
  \frac{dN}{dE} = K \left(\frac{E}{E_0}\right)^{-\Gamma} \exp\,\left(-a \left(\frac{E}{1\,{\rm MeV}}\right)^{b}\right)\,,
\end{equation}
where the parameters $E_0 = 1$\,GeV (``pivot energy'') and $b = 2/3$
(exponential index) were fixed at their \citetalias{4FGL} values, while the
parameters $K$ (normalisation), $\Gamma$ (low-energy spectral index) and $a$
(exponential factor) were free to vary during fitting. We performed a binned
likelihood analysis using \texttt{fermipy} \citep{fermipy} version 0.18.0, with
$0.05\degree\times0.05\degree$ bins and 10 logarithmic energy bins per
decade. For this analysis we utilised the ``PSF'' event types and corresponding
IRFs, which partition the LAT data into quartiles based on the quality of the
reconstructed photon arrival directions. All \citetalias{4FGL} sources within
$25\degree$ of the optical counterpart position were included in the
model. Using the ``optimize'' function of \texttt{fermipy}, the parameters of
all sources in the region were updated from their \citetalias{4FGL} values one
at a time to find a good starting point. We then performed a full fit for the
region surrounding J2039. The spectral parameters of all sources within
$10\degree$ were free to vary in the fitting, as were the normalisations of the
diffuse models and the spectral index of the Galactic diffuse model.

The gamma-ray source at the location of the optical counterpart is detected with
test statistic $\textrm{TS} = 2167$ (the TS is defined as twice the increase
in log-likelihood when the source is added to the model). The spectrum has a
photon power-law index of $\Gamma = 1.4 \pm 0.1$ and an exponential factor of
$a = (7 \pm 1)\times10^{-3}$. The total energy flux above $100$~MeV is
$G_{\gamma} = (1.46 \pm 0.06) \times 10^{-11}$~erg~cm$^{-2}$~s$^{-1}$. At an
assumed distance of $1.7$~kpc (from our optical light-curve modelling in
Section~\ref{s:modelling}), this gives a gamma-ray luminosity of $L_{\gamma} =
(5.0\pm0.6)\times10^{33}$~erg~s$^{-1}$, assuming isotropic emission.

In gamma-ray pulsation analyses, photon weights are used to weight the
contribution of each photon to a pulsation detection statistic to increase its
sensitivity, and avoid the need for hard cuts on photon energy and incidence
angle \citep{Kerr2011}. A weight $w_j$ represents the probability that the
$j$-th photon was emitted by a target source, as opposed to by a fore/background
source, based on the reconstructed photon energy and arrival direction, and a
model for gamma-ray sources within the RoI. We computed these weights for
photons whose arrival directions were within $5\degr$ of J2039 using
\texttt{gtsrcprob}, again making use of the PSF event types. Within this region,
there were $181,813$ photons in total, with $\sum_j w_j = 3850$ ``effective''
photons. To speed up our timing analyses (Section~\ref{s:timing}) we
additionally removed photons with $w < 0.1$, leaving $6571$ photons which
account for $93$\% of the expected pulsation signal-to-noise ratio \citep[which is
  proportional to $\sum_j w_j^2$, ][]{Clark2017+FGRP4}.

The data set described above was used for the timing (Section~\ref{s:timing})
and orbital modulation analyses (Section~\ref{s:gamma_var}) presented in this
paper. For the pulsation search described in Section~\ref{s:pulsar}, we used an
earlier data set which only covered data up to 2019 January 10 and used spectral
parameters from a preliminary
version\footnote{\url{https://fermi.gsfc.nasa.gov/ssc/data/access/lat/fl8y/}} of
the \citetalias{4FGL} catalogue when computing photon weights.

\subsection{Gamma-ray Pulsation Search}
\label{s:pulsar}
Using the hierarchical search methods described by \citet{Pletsch2014+Methods},
extended to provide sensitivity to binary pulsars by \citet{Nieder2020+Methods}, we performed a search for gamma-ray pulsations in the weighted
\textit{Fermi}-LAT photon arrival times.

For this, it was necessary to search for an unknown spin frequency $\nu$,
spin-down rate $\dot{\nu}$, as well as the orbital period $P_{\rm orb}$,
pulsar's time of ascending node $T_{\rm asc}$, and pulsar's projected semi-major
axis $x = A_{\rm psr} \sin i$, where $A_{\rm psr}$ is the (non-projected)
semi-major axis, and $i$ is the binary inclination angle. We did not search a range of sky positions as we used the precise \textit{Gaia} position of the optical
counterpart.

This 5-dimensional parameter volume is extremely large, and requires large
computing resources and efficient algorithms to cover. To meet the large
computational cost of the searches, we utilised the distributed volunteer
computing system, \textit{Einstein@Home}
\citep{Knispel2010+EAH,Allen2013+EAH}. Under this system, the parameter space is
split into millions of smaller chunks which can be searched by a typical
personal computer within a few hours. These ``work units'' are computed while
volunteer's computers are otherwise idle. We also ported our
\textit{Einstein@Home} search code from CPUs to GPUs, which has previously been
done for radio pulsar searches \citep{Allen2013+EAH}. The approximately 10,000
GPUs active on \textit{Einstein@Home} increase the computing speed by an order
of magnitude.

Despite this large computational resource, major efficiency gains and
compromises are required to ensure that the computational effort of the search
remains feasible. Key to improved efficiency is ensuring that the parameter space is
covered by a grid of search locations that is as sparse as possible, yet
sufficiently covers the volume to avoid missing signals. The required density is
described by a \textit{distance metric} -- a function relating parameter space
offsets to a corresponding expected loss in signal strength. This metric is described by \citet{Nieder2020+Methods}.

In the binary pulsar search, the spin parameters are searched in the same way as
they are in isolated pulsar searches \citep[see,
  e.g.,][]{Clark2017+FGRP4}. $\nu$ is searched via Fast Fourier Transforms
(FFTs). The relevant range in $\dot{\nu}$ is covered by a frequency-independent
lattice.

The computational effort to search the orbital parameters depends linearly on the
number of grid points. Searching the orbital parameters in a uniformly-spaced
grid would be inefficient because the required metric spacing depends strongly
on $\nu$ and $x$, i.e. at higher values for $\nu$ and $x$ the grid needs to be
denser \citep{Nieder2020+Methods}. To deal with the $\nu$-dependency, we
break down the search into discrete $8$\,Hz bands which are searched separately,
and in each band the grid over the orbital parameters is designed to be dense
enough for the maximum frequency in the band.

The orbital grid would be \textit{optimal} if it has the lowest number of grid
points such that each point in the parameter space is ``covered''. A location in
the parameter space is covered if the distance to the closest grid point is less
than a chosen maximum. In inhomogeneous parameter spaces, the optimal grid is
unknown. However, the required number of grid points $N_{\rm opt}$ for such a
grid can be estimated using the distance metric under the assumption that
locally the parameter space is sufficiently flat.

To search the inhomogeneous ($x$-dependent) orbital-parameter space efficiently,
\textit{optimised grids} are used \citep{Fehrmann2014}. These are built from
\textit{stochastic grids}, which are grids where grid points are placed
stochastically while no two grid points are allowed to be closer than a minimum
distance \citep{Harry2009+StochasticGrids}. We create a stochastic grid with
$N_{\rm opt}$ grid points and optimise it by nudging the position of each grid
point one by one towards ``uncovered space'' using a neighbouring cell algorithm
\citep{Fehrmann2014}. After a few nudging iterations over all grid points the
covering is typically sufficient for the search.
	
Using preliminary results from our optical modelling (see
Section~\ref{s:modelling}), obtained prior to the publication by
\citet{Strader2019+RBSpec} of spectroscopic radial velocities which better
constrain $T_{\rm asc}$, we constrained our orbital search space to $P_{\rm orb}
= 0.2279799(3)\,$d and $T_{\rm asc} = \textrm{MJD}~56884.9678(8)$. The range of expected
$x$ values was not well constrained by this model, and as the computing cost
increases with $x^3$ we chose to initially search up to $x=0.5$\,lt-s, with the
intention of searching to higher values should the search be unsuccessful.
	
The search revealed a signal with $\nu \approx 377$\,Hz that was highly
significant in both the initial semi-coherent and fully coherent follow-up
search stages. The signal had $x \approx 0.47$\,lt-s, which along with the
companion's radial velocity measurements by \citet{Strader2019+RBSpec} gives a
mass ratio of $q = M_{\rm psr}/M_{\rm c} = K_{\rm c} P_{\rm orb} / 2\pi x \approx
7.2$, and a minimum companion mass of $M_{\rm c} > 0.15\,M_{\odot}$ assuming $i
= 90 \degr$. These features conclusively confirm that the source is indeed a
redback millisecond pulsar system, which can now be named PSR~J2039$-$5617.

\subsection{Gamma-ray Timing}
\label{s:timing}
Following the discovery of gamma-ray pulsations, we used the \textit{Fermi}-LAT
data set to obtain a rotational ephemeris spanning 11 years. To do so, we followed 
the principles described by \citet{Kerr2015+FermiTiming}, in which a template pulse
profile $F(\phi)$ is produced, and the parameters $\boldsymbol{\lambda}$ of
a phase model $\phi(t \,|\, \boldsymbol{\lambda})$, are fit to
maximise the Poisson log-likelihood of the unbinned photon phases. Assuming
that the weights derived in Section~\ref{s:fermi} represent the probability that
each photon was emitted by the pulsar, then the contribution to the pulsation
log-likelihood from the $j$-th photon, with weight $w_j$, is a mixture
model between a constant (i.e. uniform in phase) background rate and the
template pulse profile, with mixture weights of $1-w_j$ and $w_j$
respectively. Hence, the overall log-likelihood is
\begin{equation}
\log L(\boldsymbol{\lambda}\,|\,t_j,w_j,F) = \sum_{j=1}^N \log \left[w_j F\left(\phi\left(t_j\,|\,\boldsymbol{\lambda}\right)\right) + (1 - w_j)\right]
\label{e:logL}
\end{equation}
where $t_j$ denotes the measured arrival time of the $j$-th detected gamma-ray
photon.

Folding the LAT data with the initial discovery ephemeris showed that the signal
was not phase-connected over the entire data span, with the pulse profile
drifting in and out of focus, indicative of a varying orbital period. Such
effects are common among redback pulsars, and are attributed to variations in
the quadrupole moment of the companion star coupling with the orbital angular
momentum \citep[e.g.,
][]{Arzoumanian1994+B1957_OPV,Lazaridis2011+J2051,Pletsch2015+J2339}. These effects significantly complicate efforts to time redbacks over more than a few months \citep[e.g.,][]{Deneva2016+J1048}.

In previous works, these
effects have been accounted for by adding a Taylor series expansion of the
orbital frequency perturbations to the constant-period orbital phase model,
where the derivatives of the orbital angular frequency become additional
parameters in the timing model. However, this parameterisation has a number of
drawbacks. Large correlations between the orbital frequency derivatives greatly
increase the time required for a Markov Chain Monte Carlo (MCMC) sampling
procedure, which suffers from inefficient sampling and exploration during the
``burn-in'' phase for highly correlated parameter spaces. The Taylor series
model also has poor predictive power as the orbital phase model ``blows up''
when extrapolating beyond the fit interval, making it difficult to extend an
existing timing solution to incorporate new data. An astrophysical
interpretation of the resulting timing solution is also not straightforward, as
the measured orbital frequency derivatives depend on one's choice of reference
epoch ($T_{\rm asc}$), and are not representative of long-term trends in $P_{\rm orb}$ due to e.g. mass loss from the system.

These problems are very similar to those encountered when timing young pulsars
with strong ``timing noise'': unpredictable variations in the spin frequency
over time. To address these issues, modern timing analyses treat timing noise as
a stationary noise process, i.e. a random process with a constant correlation function, on
top of the long-term spin-down due to the pulsar's braking
\citep{Coles2011+TimingNoise}.

\citet{Kerr2015+FermiTiming} used this method to time gamma-ray pulsars using
\textit{Fermi}-LAT data. To do this, a template pulse profile is constructed and
cross correlated with the photon phases within weeks- or months-long segments to
obtain a discrete pulse phase measurement, or ``time-of-arrival'' (TOA), for
each segment, and the stochastic noise process is fit to these phase
measurements. Timing parameters can then be fit analytically to minimise the
chi-square log-likelihood of the covariance-transformed TOA residuals including
a Bayesian penalty factor for the required timing noise process. However, this
procedure has the drawback that for faint pulsars the segment length required to
obtain a significant TOA measurement can become very long, and phase variations
due to timing noise \textit{within} each segment can no longer be neglected. Of
course, the timing noise within a segment cannot be accounted for without a
description of the noise process, which in turn cannot be obtained without the
TOAs, creating a circular problem.

While this circular problem can be partially overcome by fitting iteratively, we have
developed a new method to fit the noise process using every individual photon,
rather than obtaining and fitting discrete TOAs. To obtain this best-fitting
function and its uncertainty, we apply the \textit{sparse online Gaussian
  process} (SOGP) procedure developed by \citet{Csato2002+SOGP}. For purely
Gaussian likelihoods, the Gaussian process framework would allow an exact
posterior distribution for the noise process to be computed analytically
\citep{Rasmussen+GPML}. In our case, however, the likelihood for each photon
phase in Equation~(\ref{e:logL}) is a mixture model of Gaussian peaks describing the template pulse profile with a constant
background level. \citet{Seiferth2017+GPMM} describe how to apply the SOGP
procedure to obtain an optimal Gaussian approximation to the posterior
distribution for a stationary process with Gaussian mixture model likelihoods,
and we use this formulation to derive our timing solution. 

For J2039, we require a timing model which accounts for variations in the
orbital phase, which we treat as a stationary random process. The overall goal
is therefore to find the best-fitting continuous function describing the phase deviations from a constant orbital period model, given a
prior covariance function ($C_0(t_1,t_2)$).

Before fitting, we must choose the form of the prior covariance function, and
hyperparameters controlling its properties. Here we assumed a Mat\'{e}rn covariance function \citep{Rasmussen+GPML},
\begin{equation}
  C_0(t_1,t_2) = \frac{h^2 \,2^{1-n}}{\Gamma(n)} \left(\frac{\sqrt{2n}}{\ell}\left|t_1 - t_2\right|\right)^n K_n \left(\frac{\sqrt{2n}}{\ell}\left|t_1-t_2\right|\right)\,,
  \label{e:matern}
\end{equation}
where $K_n$ is the modified Bessel function. The hyperparameters are the length
scale, $\ell$, controlling the time span over which the orbital period remains
correlated, an amplitude parameter, $h$, which describes the expected magnitude
of the orbital phase variations, and the degree $n$, which controls the
smoothness of the noise process. In the limit of $n \to \infty$, this reduces to
the simpler squared-exponential covariance function,
\begin{equation}
  C_0(t_1,t_2) = h^2 \, \exp \left(-\frac{\left|t_1 - t_2\right|^2}{2\ell^2}\right)\,.
  \label{e:sqexp_cov}
\end{equation}

In the frequency domain, a noise process with the Mat\'{e}rn covariance function
of Equation~(\ref{e:matern}) has a power spectral density,
\begin{equation}
  P(f) \propto h^2\left(1 + \left(\frac{f}{f_{\rm c}}\right)^2\right)^{-(n + 1/2)},
  \label{e:cov_psd}
\end{equation}
i.e. constant below a ``corner frequency'' of $f_{\rm c} =
\sqrt{n}/\sqrt{2}\pi\ell$, and breaking smoothly to a power-law process with index $-(2n +
1)$ at higher frequencies.

With our chosen covariance function, we obtain a timing solution by varying the
timing parameters $\boldsymbol{\lambda}$ and hyperparameters $(\ell,h,n)$ using
the \texttt{emcee} Markov-Chain Monte Carlo (MCMC) algorithm \citep{emcee}. At
each MCMC sample, we use the \texttt{PINT} software package
\citep{Luo2018+PINT} to phase-fold the gamma-ray data according to the timing
parameters, and then apply the SOGP method to find the best-fitting Gaussian
approximation to the posterior distribution of the continuous function
describing the orbital phase variations. This posterior is marginalised
analytically, and the log marginal likelihood passed to the MCMC algorithm. This
allows the MCMC process to optimise both the timing parameters and the
hyperparameters of the prior covariance function simultaneously.

\begin{figure*}
	\includegraphics[width=0.9\textwidth]{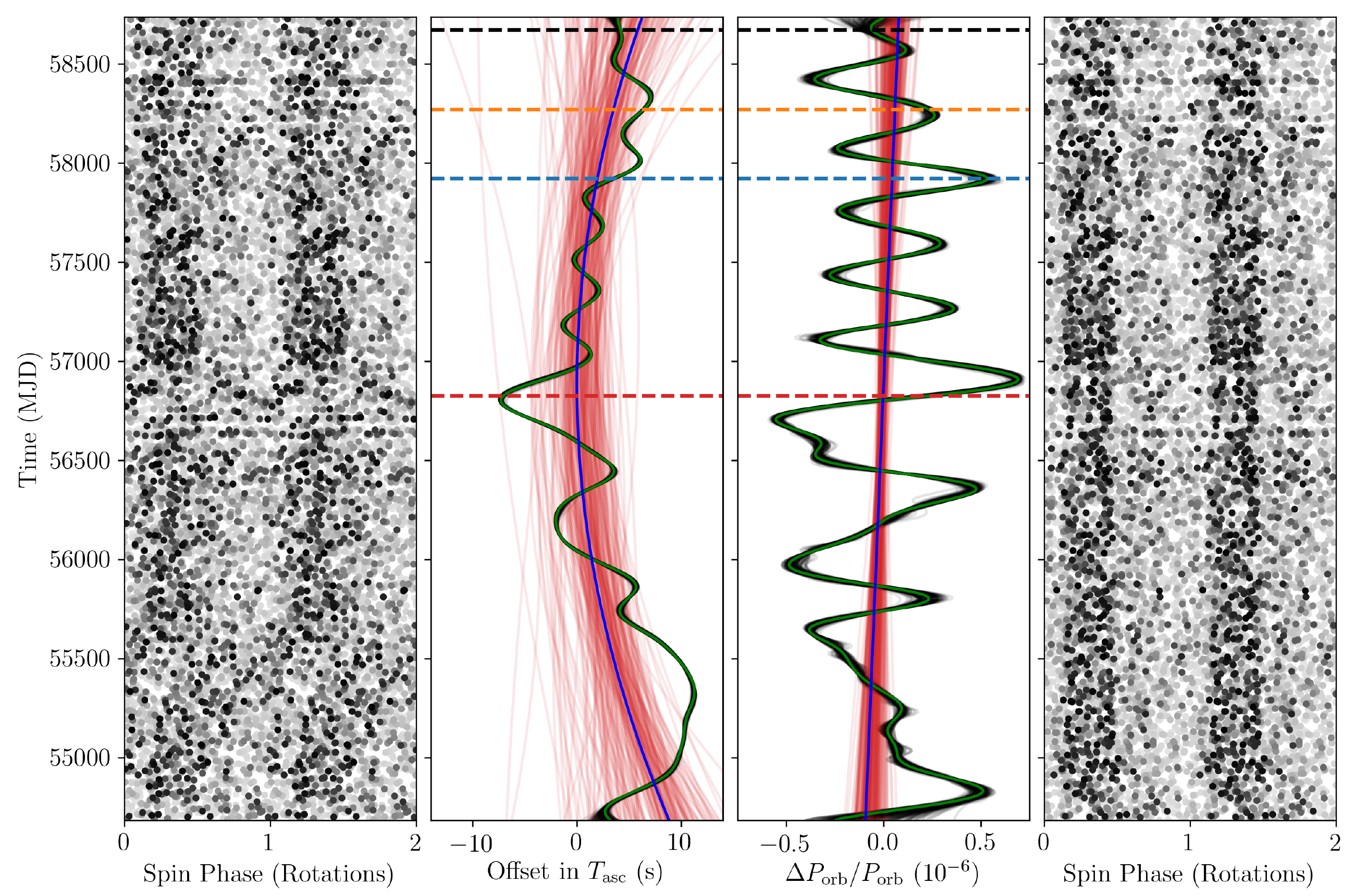}
  \caption{Results from gamma-ray timing analysis. Left panel: photon phases
    after folding with the original discovery ephemeris (with a constant orbital
    period). The intensity of each point represents the corresponding
    probability weight for that photon. The apparent loss of signal around MJDs
    55500 and 56800 is due to the varying orbital period. Although present
    throughout the entire data set, the deviations between the true orbital
    phase and that predicted by the constant-orbital-period folding model are at
    their largest at these epochs. Centre left panel: offset in the time of the
    pulsar's ascending node from the initial constant orbital period
    ephemeris. In the timing procedure we fit for an ``average'' orbital phase,
    period and first frequency derivative, and model the orbital phase
    variations as a Gaussian process on top of this base model. Variations
    requiring a Gaussian process with a larger amplitude or more complexity suffer
    a Bayesian penalty factor. Black and red lines show the best-fitting orbital
    phase variations and the underlying ``average'' orbital model, respectively,
    for randomly selected samples from the MCMC process. Green and blue curves
    show the samples with the highest log marginal likelihood. The epochs of our
    optical observations are marked by horizontal dashed lines with the same
    colour as the corresponding light curves in Section~\ref{s:optical}. Centre right
    panel: as before but for the orbital period (i.e. derivatives of the curves
    in the previous panel). Right panel: photon phases after correcting for the
    orbital phase variations using the best-fitting parameter values. }
    \label{f:OPV}
\end{figure*}

Using the best-fitting timing solution, we then re-fold the photon arrival
times, and update the template pulse profile. This process is applied
iteratively until the timing parameters and template pulse profile
converges. For J2039, this required three iterations. The results from our
timing analyses of J2039 are shown in Figure~\ref{f:OPV}
and the resulting parameter estimates are given in Table~\ref{t:params}.

\begin{figure}
  \includegraphics[width=\columnwidth]{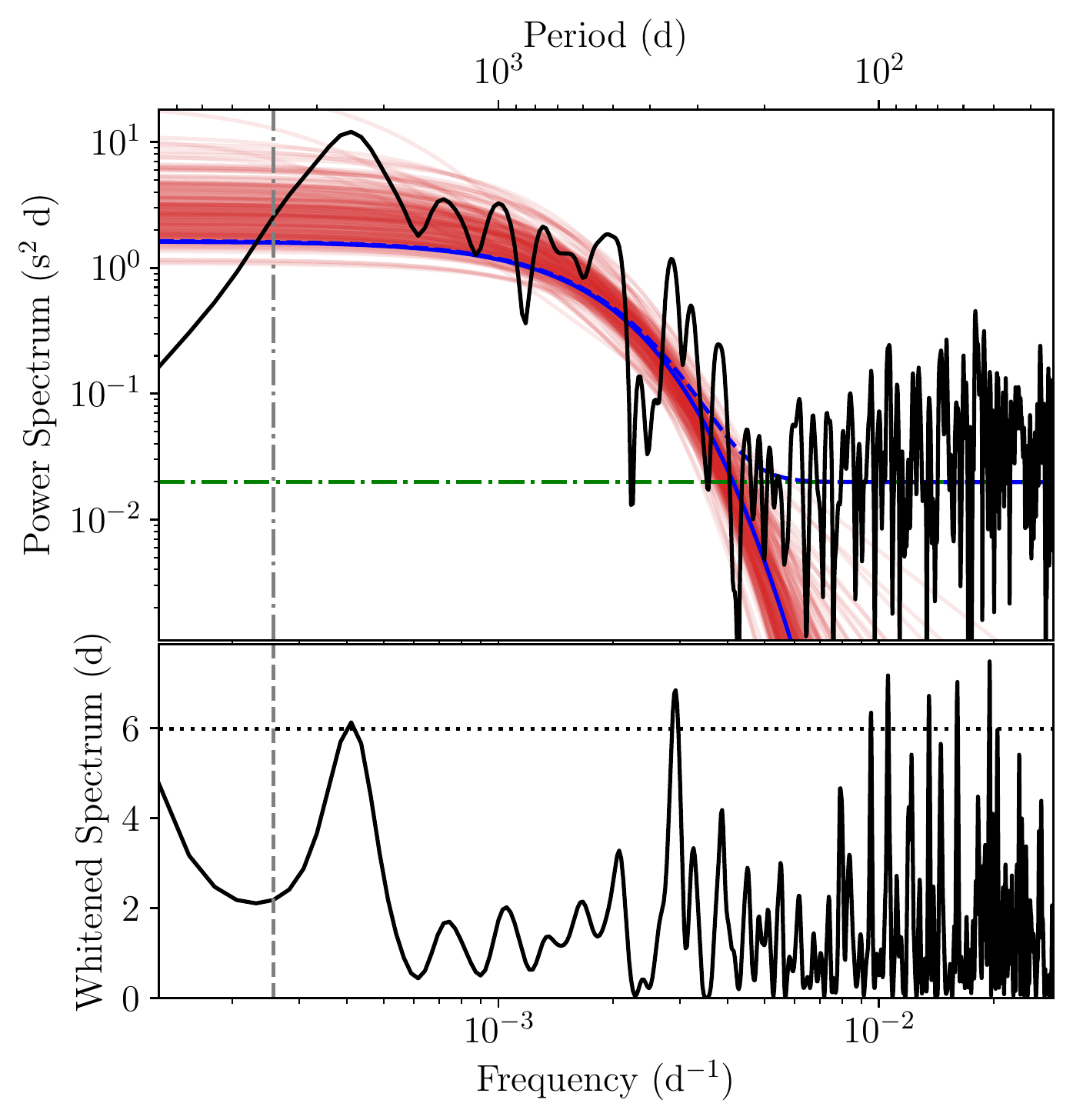}
  \caption{ \label{f:opv_psd} Power spectral density of the orbital phase noise
    process. The top panel shows the power spectral densities for the orbital
    phase variations. The green horizontal dot-dashed line shows the estimated
    measurement uncertainty level, to which the power spectrum breaks at high
    frequencies. The solid blue and red curves show the best-fitting Mat\'{e}rn
    covariance function model, and those of random samples from the MCMC
    process, respectively. The dashed blue line additionally includes the
    measurement noise level. The lower panels show the power spectra of the
    orbital phase residuals after whitening using the Cholesky decomposition of
    the model covariance matrix (i.e. accounting for the blue curve in the upper
    panel). The horizontal dotted line shows the estimated level which the noise
    power in $95$\% of independent trials should be below. The vertical line in
    both panels shows the time span covered by the \textit{Fermi}-LAT data --
    noise power close to and below this frequency is suppressed by our inclusion
    of a single orbital period derivative in the timing model. }
\end{figure}

We also show the amplitude spectra of the orbital phase variations and our best
fitting covariance model in Figure~\ref{f:opv_psd}. This spectrum was
estimated by measuring the orbital phase in discrete segments of data, and
performing the Cholesky least-squares spectral estimation method of
\citet{Coles2011+TimingNoise}. This is only used to illustrate the
later discussion (Section~\ref{s:opv}), while statements about the measured
hyperparameter values are from the full unbinned timing procedure described
above.

We have extended \texttt{TEMPO2} \citep{Edwards2006+Tempo2} with a function that
interpolates orbital phase variations between those specified at user-defined
epochs. This allows gamma-ray or radio data to be phase-folded using the
ephemerides that result from our Gaussian process model for orbital period
variations.

\begin{table}
	\centering
	\caption{Timing solution for PSR~J2039$-$5617. Epochs and units are in
          the Barycentric Dynamical Time (TDB) system.  The numerical values for
          timing parameters are the mean values of the MCMC samples, with
          $1\sigma$ uncertainties on the final digits quoted in brackets.}
	\label{t:params}
	\begin{tabular}{lc}
		\hline
    Parameter & Value\\
		\hline
    \hline
    \multicolumn{2}{c}{Astrometric Parameters$^{\rm a}$}\\
    \hline
    R.A. (J2000), $\alpha$ & $20^h39^m34\fs9681(1)$\\
    Decl. (J2000), $\delta$ & $-56\degr17\arcmin09\farcs268(1)$\\
    Proper motion in R.A., $\mu_{\alpha}\cos\delta$ (mas yr$^{-1}$) & $4.2(3)$\\
    Proper motion in Decl., $\mu_{\delta}$ (mas yr$^{-1}$) & $-14.9(3)$\\
    Parallax, $\varpi$ (mas) & $0.40(23)$\\
    Position reference epoch (MJD) & $57205.875$\\
    \hline
    \multicolumn{2}{c}{Timing Parameters}\\
    \hline
    Solar System Ephemeris & DE430\\
    Data span (MJD) & $54682$--$58738$\\
    Spin frequency reference epoch, $t_{\rm ref}$ (MJD) & 56100\\
    Spin frequency, $\nu$ (Hz) & $377.22936337986(5)$\\
    Spin-down rate, $\dot{\nu}$ (Hz s$^{-1}$) & $-2.0155(6)\times10^{-15}$\\
    Spin period, $P$ (ms) & $2.6509071060648(5)$\\
    Spin period derivative, $\dot{P}$ & $1.4164(4)\times10^{-20}$\\
    Pulsar's semi-major axis, $x$ (lt s) & $0.47105(1)$\\
    Epoch of pulsar's ascending node, $T_{\rm asc}$ (MJD) & $56884.96698(2)$\\
    Orbital period, $P_{\rm orb}$ (d) & $0.227979805(3)$\\
    Orbital period derivative, $\dot{P}_{\rm orb}$ & $8(5)\times10^{-12}$\\
    Amplitude of orbital phase noise$^{\rm b}$, $h$ (s) & $3.9^{+2.2}_{-1.1}$\\
    Correlation timescale$^{\rm b}$, $\ell$ (d) & $156^{+127}_{-41}$\\
    Mat\'{e}rn function degree$^{\rm c}$, $n$ & $>1.5$\\
    \hline
    \multicolumn{2}{c}{Derived properties$^{\rm d}$}\\
    \hline
    Shklovksii spin down, $\dot{\nu}_{\rm Shk}$ (Hz s$^{-1}$) & $(-0.37 \pm 0.02)\times10^{-15}$\\
    Galactic acceleration spin down, $\dot{\nu}_{\rm acc}$ (Hz s$^{-1}$) & $1.2 \times10^{-17}$\\
    Spin-down power, $\dot{E}$ (erg s$^{-1}$) & $2.5\times10^{34}$\\
    Surface magnetic field strength, $B_{\rm S}$ (G) & $2\times10^{8}$\\
    Light cylinder magnetic field strength, $B_{\rm LC}$ (G) & $8.8\times10^{4}$\\
    Characteristic age, $\tau_{\rm c}$ (yr) & $4\times10^{9}$\\
    Gamma-ray luminosity, $L_{\gamma}$ (erg s$^{-1}$) & $(5.0 \pm 0.6) \times10^{33}$\\
    Gamma-ray efficiency, $\eta_{\gamma} = L_{\gamma} / \dot{E}$ & $0.21$\\
    \hline
    \multicolumn{2}{p{\columnwidth}}{\scriptsize $^{\rm a}$ Astrometric parameters are taken from \citet{GAIA2018+DR2}. }\\
    \multicolumn{2}{p{\columnwidth}}{\scriptsize $^{\rm b}$ The hyperparameters $h$ and $\ell$ have asymmetric posterior distributions, and so we report the mean value and 95\% confidence interval limits in super- and subscripts.}\\
    \multicolumn{2}{p{\columnwidth}}{\scriptsize $^{\rm c}$ The Mat\'{e}rn function degree $n$ is poorly constrained by the data; we report only a 95\% confidence lower limit.}\\
    \multicolumn{2}{p{\columnwidth}}{\scriptsize $^{\rm d}$ Derived properties are order-of-magnitude estimates calculated using the following expressions \citep[e.g.,][]{2PC}, which assume a dipolar magnetic field, and canonical values for the neutron-star moment of inertia, $I = 10^{45}\,\textrm{g cm}^2$ and radius, $R = 10$\,km: $\dot{E} = -4 \pi^2 I \nu \dot{\nu}$; $B_{\rm S} = \sqrt{-1.5 I c^3 \dot{\nu} \nu^{-3}} / (2 \pi R^3)$; $B_{\rm LC} = 4\pi^2 \sqrt{-I \dot{\nu} \nu^3 / c^3}$; $\tau_{\rm c} = \nu / 2 \dot{\nu}$. The corrections to $\dot{\nu}$ due to transverse motion (the Shklovskii effect) and radial acceleration in the Galactic potential were applied prior to computing other derived properties, assuming $d=1.7$\,kpc from optical light curve modelling described in Section~\ref{s:modelling}. }\\
  \end{tabular}
\end{table}

\subsection{Gamma-ray Variability}
\label{s:gamma_var}
The subset of transitional redback systems has been seen to transition to and
from long-lasting accretion-powered states, in which their gamma-ray flux is
significantly enhanced
\citep{Stappers2014+J1023,Johnson2015+J1227,Torres2017+tMSPs}. To check for such
behaviour from J2039, we investigated potential gamma-ray variability over the
course of the \textit{Fermi}-LAT data span. In \citetalias{4FGL}, J2039 has two-month and
one-year variability indices (chi-squared variability tests applied to the
gamma-ray flux measured in discrete time intervals) of $44$ with $48$ degrees of
freedom, and $13$ with $7$ degrees of freedom, respectively. Although the 1-year
variability index is slightly higher than expected for a steady source, we note
that the gamma-ray light curves in \citet{Ng2018+J2039_orb} indicate that a
flare from a nearby variable blazar candidate, 4FGL~J2052.2$-$5533, may have
contaminated the estimated flux from J2039 around MJD 57100. The true
variability is therefore likely lower than suggested by the slightly elevated
annual variability index, and indeed the two-month variability index is
consistent with a non-variable source.

We also checked for a potential gamma-ray eclipse, which may occur if the binary
inclination angle is high enough that the pulsar passes behind the companion
star around superior conjunction, as has been observed in the transitional MSP
candidate 4FGL~J0427.8$-$6704
\citep[][]{Strader2016+J0427,Kennedy2020+J0427}. For J2039, this would occur for
inclinations $i\gtrsim78\degr$, and could last for up to $7$\% of an orbital
period, assuming a Roche-lobe filling companion. We modelled the eclipse as a
simple ``top-hat'' function, in which the flux drops to zero within the eclipse,
and used the methods described by \citet{Kerr2019+godot}, and applied to the
eclipse of 4FGL~J0427.8$-$6704 by \citet{Kennedy2020+J0427}, to evaluate the log-likelihood of this model given the observed
photon orbital phases. We find that an eclipse lasting longer than $0.1$\% of an
orbit is ruled out by the gamma-ray data with $95$\% confidence. We interpret
this as evidence that the pulsar is not eclipsed, and will use this to constrain
the binary inclination while modelling the optical light curves in
Section~\ref{s:modelling}.

\subsubsection{Gamma-ray Orbital Modulation}
\label{s:orbvar}
As noted previously, \citet{Ng2018+J2039_orb} discovered an orbitally modulated
component in the gamma-ray flux from 4FGL~J2039.5$-$5617. Using the now
precisely determined gamma-ray timing ephemeris (see Section~\ref{s:timing}) we
computed the orbital Fourier power of the weighted photon arrival times, finding
$P=29.7$ for a slightly more significant single-trial false-alarm probability of
$p_{\rm FA} = \textrm{e}^{-P/2} \approx 4\times10^{-7}$ compared to that found
by \citet{Ng2018+J2039_orb}. Those authors found the modulation was not detected
after MJD 57040 and speculated that this could be due to changes in the relative
strengths of the pulsar wind and companion wind/magnetosphere. We do see a slight leveling-off in
the rate of increase of $P$ with time; however it picks up again after MJD
58100. Variations in the slope of this function due to statistical (Poisson)
fluctuations can appear large when the overall detection significance is low
\citep{Smith2019+1kPSRs}, and so we do not consider this to be compelling
evidence for long-term flux variability from the system.

\begin{figure}
  \includegraphics[width=\columnwidth]{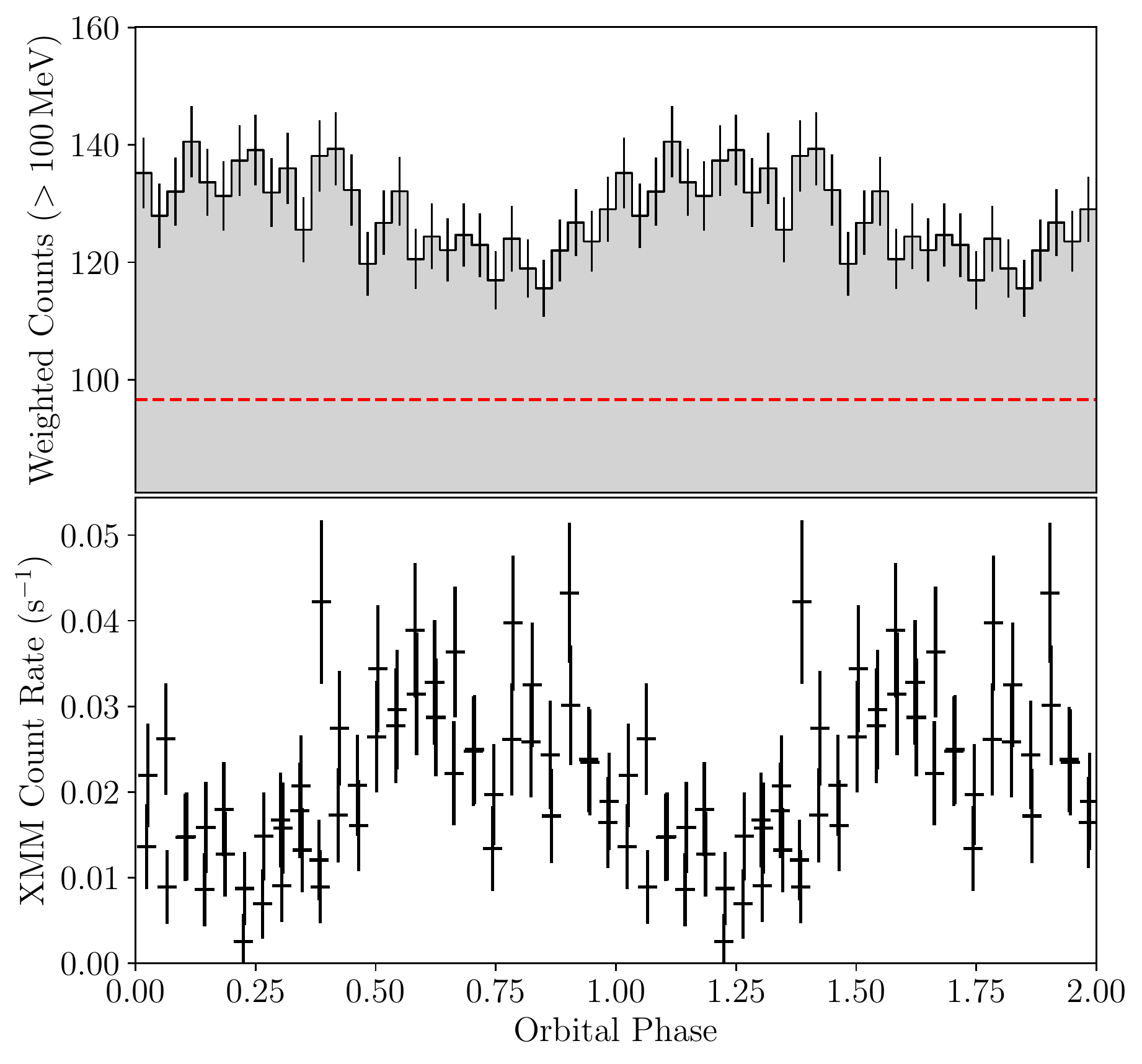}
  \caption{  \label{f:Fermi_XMM_Folded}
    Orbital light curves of J2039 from \textit{XMM-Newton} (lower panel)
    and \textit{Fermi}-LAT (upper panel) observations. Data have been folded
    using the pulsar timing ephemeris from Section \ref{s:timing}. The
    dashed red horizontal line on the gamma-ray light curve indicates the
    expected background level computed from the distribution of photon weights.}
\end{figure}

The gamma-ray and X-ray orbital light curves are shown in
Figure~\ref{f:Fermi_XMM_Folded}.  We also find no power at higher harmonics of
the orbital period, indicating an essentially sinusoidal profile. The gamma-ray
flux peaks at orbital phase $\Phi = 0.25\pm0.03$ (pulsar superior conjunction),
almost exactly half an orbit away from the X-ray peak, and has an
energy-averaged pulsed fraction of $24\pm5$\% \citep[using the definition from
  Equation (14) of ][]{Clark2017+FGRP4}. As noted by \citet{Ng2018+J2039_orb},
this phasing might suggest an inverse Compton scattering (ICS) origin, as
opposed to being the high-energy tail of the population responsible for X-ray
synchrotron emission from the intra-binary shock, for example, which would be
phase-aligned with the X-ray modulation.

\begin{figure}
  \includegraphics[width=\columnwidth]{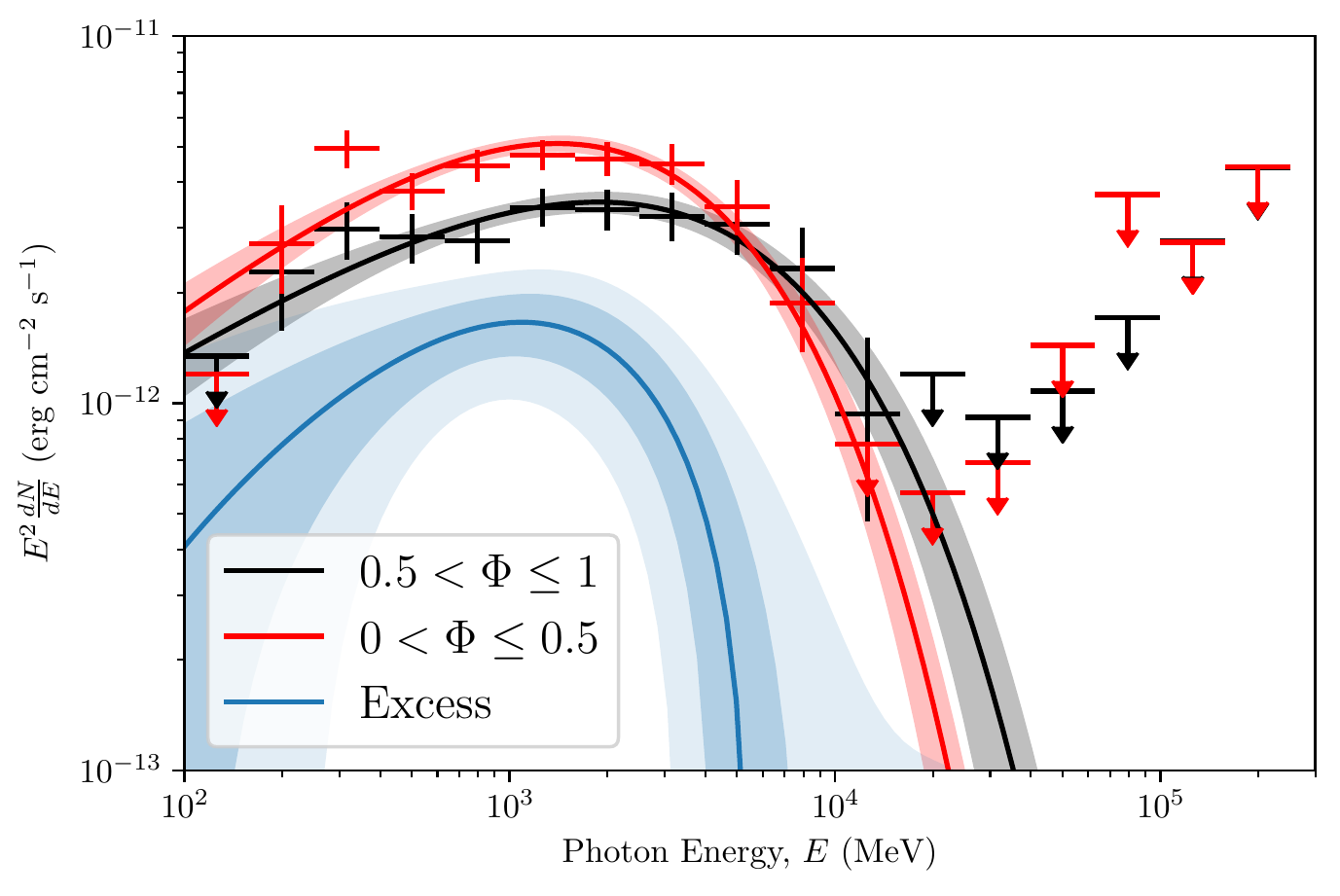}
  \caption{Gamma-ray spectral energy distributions (SEDs) for PSR~J2039$-$5617,
    measured in two discrete orbital phase ranges around pulsar superior ($0.0 <
    \Phi \leq 0.5$) and inferior ($0.5 < \Phi \leq 1.0$)
    conjunctions. Error bars are derived by fitting the normalisation of a
    power-law spectrum with index $2$ to the flux measured in five discrete
    logarithmically spaced energy bands per decade. The deviating points at low
    energies are likely due to source confusion, as seen in the SEDs of several
    sources in \citetalias{4FGL}. The curved lines and shaded regions illustrate
    the best-fitting spectral models and one-sigma uncertainties in each phase
    interval. The blue curve and shaded regions show the difference between the
    spectral models measured in the two phase intervals.  \label{f:SEDs}}
\end{figure}

To further investigate this modulation, we performed a second spectral analysis,
using the same procedure as above, but additionally separating the photons into
``maximum'' ($0.0 < \Phi \leq 0.5$) and ``minimum'' ($0.5 < \Phi_{\rm
  orb} \leq 1.0$) orbital phases. We fit the spectral parameters of J2039
separately in each component, while the parameters of other nearby sources and
of the diffuse background were not allowed to vary between the two
components. The results are given in Table~\ref{t:fermi_spectra} and the
resulting spectral energy distributions shown in Figure
\ref{f:SEDs}. Subtracting the ``minimum'' spectrum from the ``maximum''
spectrum, we find an additional component peaking at around $1\,$GeV, and
decaying quickly above that, whose total energy flux is around $30$\% of the
flux at the orbital minimum. This model has a significant log-likelihood increase of $\Delta \log L = 14$ (${\rm TS} = 28$ for a false-alarm probability of $5\times10^{-6}$ given 3 degrees of freedom) compared to our earlier model where the gamma-ray flux is constant with orbital phase.

\begin{table}
	\centering
	\caption{Gamma-ray spectral parameters in two orbital phase
          regions. Photon and energy fluxes are integrated over photon energies
          $E > 100$\,MeV. Uncertainties are at the $1\sigma$
          level. }
	\label{t:fermi_spectra}
	\begin{tabular}{lcc}
          \hline
          Parameter & $0 < \Phi \leq 0.5$ & $0.5 < \Phi \leq 1$\\
          \hline
Photon index, $\Gamma$ & $1.25\pm0.13$ & $1.42\pm0.14$ \\
Exponential factor, $a$ ($10^{-3}$) & $9.0\pm1.3$ & $5.7\pm1.2$\\
Photon flux ($10^{-8}$ cm$^{-2}$ s$^{-1}$) & $1.8\pm0.2$ & $1.3\pm0.2$\\
Energy flux, $G_{\gamma}$ ($10^{-11}$ erg cm$^{-2}$ s$^{-1}$) & $1.7\pm0.1$ & $1.3\pm0.1$\\
\hline
        \end{tabular}
\end{table}

\begin{figure}
  \includegraphics[width=\columnwidth]{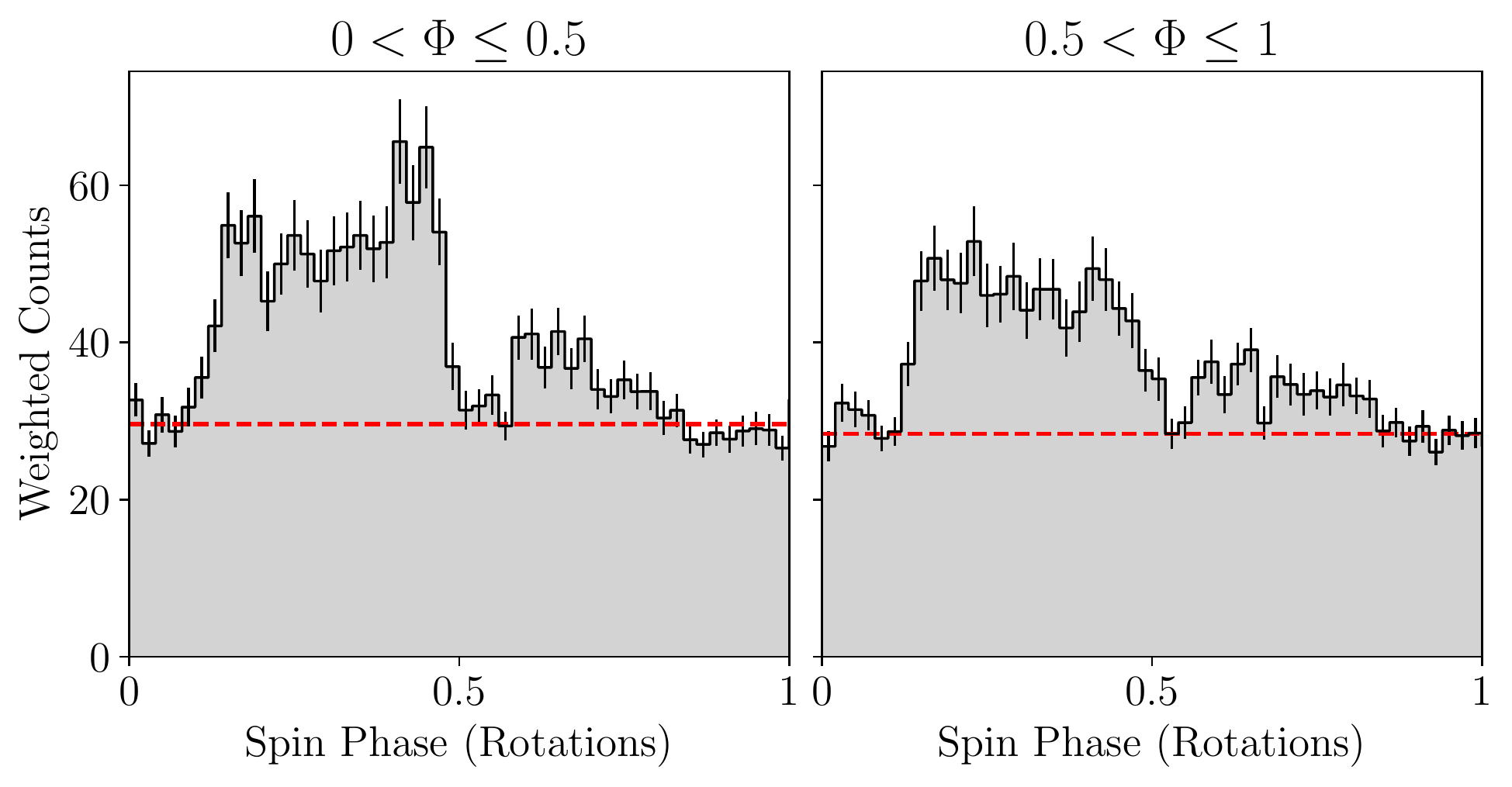}
  \caption{ \label{f:orb_mod_pulseprof} The gamma-ray pulse profile of
    PSR~J2039$-$5617 measured in data taken in two equally-sized orbital phase regions
    around the pulsar superior (left) and inferior (right) conjunctions. The red
    dashed line indicates the background level, estimated independently in each
    orbital phase region using the distribution of photon probability
    weights. The gamma-ray pulse profile is clearly enhanced around superior
    conjunction, and there is no evidence for an unpulsed component in either
    orbital phase region. }
\end{figure}

Similar orbital modulation has been observed from a handful of other spider
systems \citep{Wu2012+B1957,An2017+J1311,An2018+J2241,An2020+J2339}. In two of these systems
the gamma-ray flux peaks at the same orbital phase as is seen here from J2039,
and importantly, from the redback PSR J2339$-$0533 the orbitally modulated
component appears to be pulsed in phase with the ``normal'' intrinsic gamma-ray
pulses.

Using the timing solution from Section~\ref{s:timing}, we can now investigate
any rotational phase dependence of the orbitally-modulated component. In
Figure~\ref{f:orb_mod_pulseprof} we show the gamma-ray pulse profile, split into
two equal orbital phase regions around the pulsar superior ($0 < \Phi \leq 0.5$)
and inferior conjunctions ($0.5 < \Phi \leq 1$). We find that the estimated
background levels, calculated independently in each phase region from the photon
weights as $b = \sum_j w_j - w_j^2$ \citep{2PC}, are very similar between the
two orbital phase selections, that the pulse profile drops to the background
level in both, and that the gamma-ray pulse is significantly brighter around the
pulsar superior conjunction. There is therefore no evidence for an unpulsed
component to the gamma-ray flux from J2039, and the extra flux at the companion
inferior conjunction is in fact pulsed and in phase with the pulsar's intrinsic
pulsed gamma-ray emission. 

We consider two possible explanations for this orbitally-modulated excess. In
these models, charged particles are accelerated in an inclined, fan-like current
sheet at the magnetic equator that rotates with the pulsar. The intrinsic pulsed
gamma-ray emission is curvature radiation seen when the current sheet crosses
the line of sight. In the first scenario, the additional component is ICS from
relativistic leptons upscattering the optical photon field surrounding the
companion star. In the second, these leptons emit synchrotron radiation in the
companion's magnetosphere. These processes cause the normally unseen flux of
relativistic leptons that is beamed towards the observer when the current sheet
crosses the line of sight to become detectable as an additional pulsed gamma-ray
flux that is coherent in phase with the intrinsic emission. We shall defer a
full treatment of this additional emission component to a future work (Voisin,
G. et al. 2020, in prep), and instead discuss some broad implications of the
detection.

In the ICS scenario, it appears unlikely that the ICS population and the
population responsible for the intrinsic (curvature) emission share the same
energy. Indeed, the typical energy of the scattered photons, about $E_s \sim 1$
GeV, suggests scattering in the Thomson regime (for leptons) with $E_s \sim
\gamma_s^2 E_{b}$, where $\gamma_s$ is the typical Lorentz factor of the
scatterer and $E_b \sim 1$eV is the energy of soft photons coming from the
companion star. This implies $\gamma_s \sim 3\times 10^4$ which fulfils the
condition $E_s \ll \gamma_s mc^2$ necessary for Thomson regime scattering. On
the other hand, the Lorentz factor required to produce intrinsic gamma rays at
an energy $E_i \sim 2$ GeV is about $\gamma_i \sim 10^7$ assuming the mechanism
is curvature radiation \citep[as is favoured by][]{Kalapotharakos2019+FP}.
We assumed a curvature radius equal to the light-cylinder radius
$r_{\mathrm{LC}} = 126\,$km and a magnetic field intensity equal to
$B_{\mathrm{LC}} = 7\times10^{4}\,$G in these estimates. Thus, the ICS scenario
requires two energetically distinct populations of leptons in order to explain
the orbital enhancement. Under this interpretation, the more relativistic
curvature-emitting population would also produce an ICS component peaking around
10 TeV, which may be detectable by future ground-based Cherenkov telescopes.

The synchrotron scenario, on the other hand, allows for the possibility that the
same particle population responsible for intrinsic pulsed gamma-ray (curvature)
emission can produce the orbital flux enhancement, provided the companion
magnetic field strength is on the order of $10^3$\,G \citep{Wadiasingh2018}. The
synchrotron critical frequency in a $10^3$\,G field of the companion
magnetosphere is $\sim 1$~GeV for a Lorentz factor of $\gamma_i = 10^7$, while
the cooling timescale is about $10^{-5}$ -- $10^{-4}$\,s, i.e. leptons cool
almost immediately after crossing the shock, and phase coherence can be
maintained. Moreover, the particles are energetic enough to traverse the shock
without being greatly influenced, and would emit in less than a single
gyroperiod, so emission would likely be beamed in the same direction as the
intrinsic curvature radiation.

For the pulsed orbital modulation in PSR J2339$-$0533,
\citet{An2020+J2339} also consider an alternative scenario
in which intrinsic pulsed emission is absorbed around the pulsar's inferior
conjunction.  This model explains the softer spectrum around the maximum, as
leptons in the pulsar wind have a higher scattering cross section for low-energy
gamma rays. However, they conclude that the pair density within the pulsar wind
is far too low to provide sufficient optical depth.

\section{Optical observations and Modelling}
\label{s:optical}
\subsection{New optical observations}
We performed optical photometry of J2039 with the high-speed triple-beam CCD
camera ULTRACAM \citep{Dhillon2007+ULTRACAM} on the NTT on 2017 June 18, 2018
June 02 and 2019 July 07. The first two observations each covered just over one
full orbital period, while the third was affected by intermittent cloud cover
throughout before being interrupted by thick clouds after 70\% of an orbit had
been observed. We observed simultaneously in $u_s, g_s$ and
$i_s$\footnote{ULTRACAM uses higher-throughput versions of the SDSS filter set,
  which we refer to as {\it Super-SDSS filters}: $u_s$, $g_s$, $r_s$, $i_s$, and
  $z_s$ \citep{dhillon2018}.}, with 13\,s exposures (65\,s in $u_s$) and
negligible dead time between frames. Each image was calibrated using a bias
frame taken on the same night and a flat-field frame taken during the same
observing run.

All reduction and calibration was performed using the ULTRACAM software
pipeline\footnote{\url{http://deneb.astro.warwick.ac.uk/phsaap/software/ultracam/html/}}
(GROND images were first converted to the ULTRACAM pipeline's data
format). Instrumental magnitudes were extracted using aperture photometry, with
each star's local per-pixel background count rate being estimated from a
surrounding annulus and subtracted from the target aperture.

To calibrate the photometry, we took ULTRACAM observations of two Southern SDSS
standard fields (Smith, J.A., et al. 2007, AJ,
submitted)\footnote{\url{http://www-star.fnal.gov/}} on 2018 June 01 and 2018
June 04. The resulting zeropoints were used to calibrate the ULTRACAM
observations of J2039. Zeropoint offsets between 2017, 2018 and 2019 observations
and frame-to-frame transparency variations were corrected via ``ensemble
photometry'' \citep{Honeycutt1992+EnsemblePhot} using a set of 15 stars that
were present in all ULTRACAM and GROND images of J2039.

To calibrate the archival GROND data, we computed average magnitudes for 5
comparison stars that were covered in $g_s$ and $i_s$ by the ULTRACAM
observations, and fit for a linear colour term between the GROND and ULTRACAM
filter sets. Neither $r^\prime$ nor $z^\prime$ were covered by ULTRACAM. In
$r^\prime$ we therefore used magnitudes of 4 stars from the APASS catalogue
\citep{APASS+DR10}. No catalogues contained calibrated $z^\prime$ magnitudes for
stars within the GROND images. We therefore adopted the reference GROND
zeropoint\footnote{\url{http://www.mpe.mpg.de/~jcg/GROND/calibration.html}} in
this band. The $g^\prime$, $r^\prime$ and $i^\prime$ the GROND calibrations
agreed with these reference zeropoints to within $0.07$\,mag. As a cross-check
we derived alternative zeropoints using a set of stars in the images which have
magnitudes listed in the APASS catalogue. For both GROND and ULTRACAM the
APASS-derived zeropoints agree with the calibrations using the ULTRACAM
standard-derived zeropoints to within $0.06$\,mag in both $g_s$ and $i_s$.

\begin{figure}
	\includegraphics[width=\columnwidth]{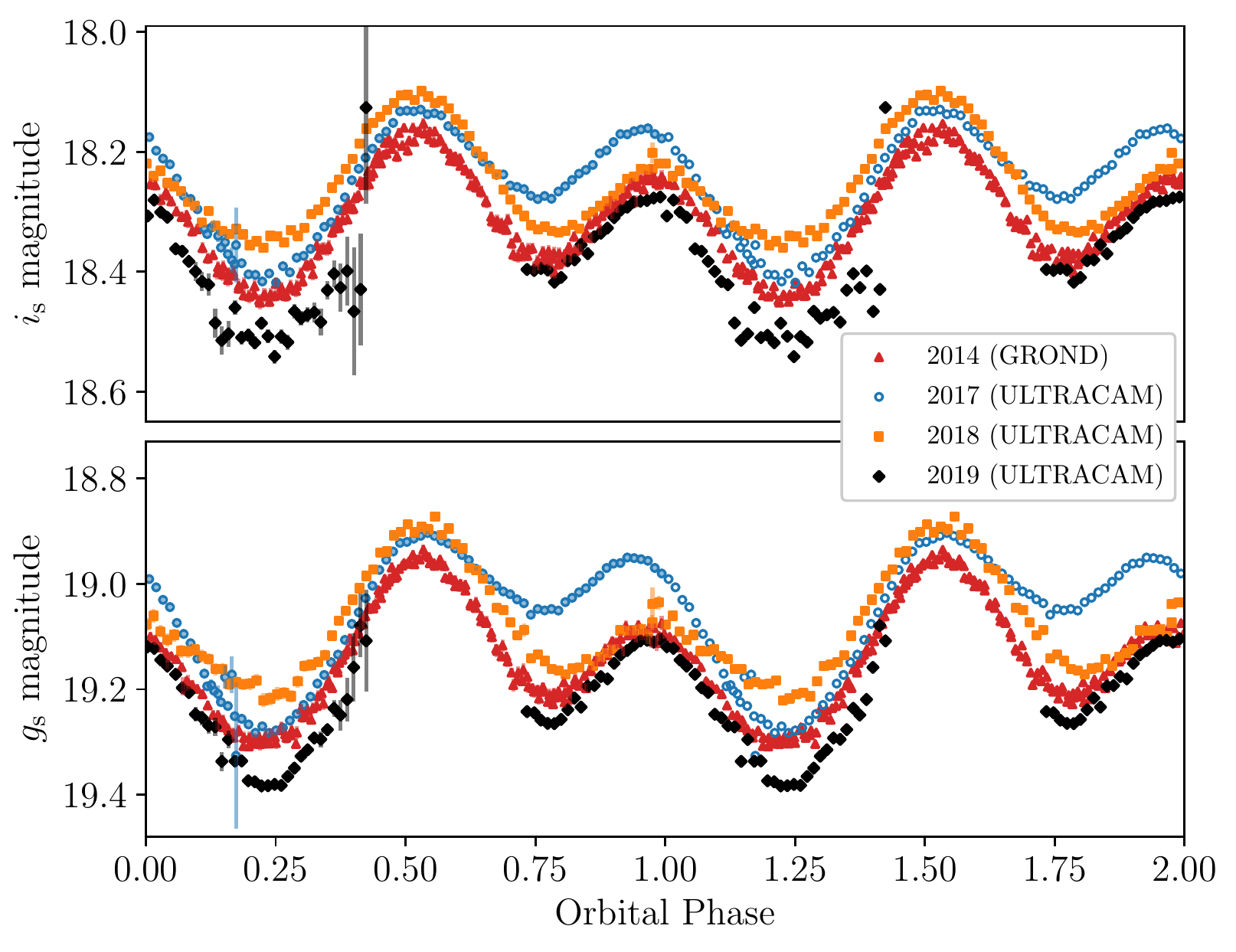}
  \caption{Folded orbital light curves of J2039 across four observing runs (2014
    August 16--18 with GROND, 2017 June 18, 2018 June 02 and 2019 July 07 with
    ULTRACAM). For clarity, the ULTRACAM data points have been combined into
    $250$\,s integrations. The folded light curves are repeated twice, with
    uncertainties shown only in the first cycle. These are mostly smaller than
    the corresponding markers. Here, and throughout this paper, orbital phase
    zero corresponds to the pulsar's ascending node. The GROND light curves have
    been corrected to the ULTRACAM magnitude system via colour corrections
    computed from the magnitudes of comparison stars in the field. }
    \label{f:LC}
\end{figure}

The resulting light curves in the $g_s$ and $i_s$ bands (the only two bands
covered by all 4 observations) are shown in Figure \ref{f:LC}. The long-term
changes in the light curve are clearly visible, with $\sim0.2$\,mag variability
in the second maximum (near the companion star's descending node) and
$\sim0.1$\,mag variations in the minimum at the companion's inferior
conjunction. The apparent variations around the first maximum (companion's
ascending node), are closer to our systematic uncertainty in the relative flux calibrations. 

To estimate the level of variability that can be attributed to our flux
calibration, we checked the recovered mean magnitudes of the ensemble stars used
to flux-calibrate the data. These all varied by less than $0.05$~mag across
all sets of observations. 

\label{s:data}

\subsection{Light curve modelling}

\label{s:modelling}
To estimate physical properties of the binary system, we fit a model of the
binary system to the observed light curves using the \texttt{Icarus} binary
light curve synthesis software \citep{Breton2012+Icarus}.

\texttt{Icarus} assumes point masses at the location of the pulsar (with mass
$M_{\rm psr}$) and companion star centre-of-masses, and solves for the size and
shape of the companion star's Roche lobe, according to an assumed binary mass
ratio $q \equiv M_{\rm psr}/M_{\rm c}$, inclination angle $i$, projected
velocity semi-amplitude $K_{\rm c}$ and orbital period $P_{\rm orb}$. These
parameters are linked through the binary mass function,
\begin{equation}
  f(M_{\rm psr}) = \frac{M_{\rm psr} \sin^3 i}{(1 + 1/q)^2} = \frac{K_{\rm c}^3
    P_{\rm orb}}{2\pi \textrm{G}}\,,
  \label{e:mass_function}
\end{equation}
and hence only 4 out of these 5 values are independent. With the pulsation
detection, we have an extremely precise timing measurement of $P_{\rm orb}$, and
the pulsar's projected semi-major axis ($x$), which further fixes $q = K_{\rm c}
P_{\rm orb} / 2\pi x$. We therefore chose to fit for $i$ and $K_{\rm c}$, and
derive $q$ and $M_{\rm psr}$ from these. For $i$, we adopted a prior that is
uniform in $\cos i$ (to ensure that the prior distribution for the orbital
angular momentum direction is uniform over the sphere). Since no evidence is
seen for a gamma-ray eclipse (see Section~\ref{s:gamma_var}), we assume that the
pulsar is not occluded by the companion star, which provides an upper limit on
the inclination of $i \lesssim 79\degr$ (the precise limit additionally depends on
the size of the star, and was computed ``on-the-fly'' by \texttt{Icarus} while
fitting). We additionally assumed a conservative lower limit of $i > 40\degr$,
since lower inclinations would require an unrealistically high pulsar mass
($>4\,M_{\odot}$).

The size and shape of the star within the Roche lobe is parameterised by the
Roche lobe filling factor $f_{\rm RL}$, defined as the ratio between the radius
from the star's centre-of-mass in the direction towards the pulsar and the
distance between the star's centre-of-mass and the Lagrange L1 point. 

Once the shape of the star has been calculated, the surface temperature of the
companion star is defined by another set of parameters. The temperature model
starts with the ``night'' side temperature of the star, $T_{\rm n}$, which is
the base temperature at the pole of the star prior to irradiation. We assumed a
Gaussian prior on $T_{\rm n}$ with mean $5423$\,K and width $\pm249$\,K taken
from the \textit{Gaia} colour--temperature relation \citep{GAIA2010+Phot}. To
account for gravity darkening, we modify the surface temperature at a given
location for the local effective gravitational acceleration by $T_{\rm g} =
T_{\rm n} (g/g_{\rm pole})^{\beta}$, where $g_{\rm pole}$ is the effective
gravitational acceleration at the pole. We used a fixed value of $\beta = 0.08$,
which assumes that the companion star has a convective envelope
\citep{Lucy1967+betaConv}.

We account for the effect of heating from the pulsar by modelling it as an
isotropically emitting point source of heating flux, with luminosity $L_{\rm
  irr}$ (although note that the pulsar's beam is generally more concentrated
towards the equator, see \citealt{Draghis2019+BWs} who account for this when
fitting black-widow light curves). In \texttt{Icarus}, heating is
parameterised by the ``irradiation temperature'' $T_{\rm irr} = L_{\rm
  irr}/(4\pi\sigma A^2)$, where $\sigma$ is the Stefan-Boltzmann constant and $A
= x (1 + q) / \sin i$ is the orbital separation. In the later discussion, we
will compare this luminosity with the pulsar's total spin-down power via the
heating efficiency, $\epsilon = L_{\rm irr} / \dot{E}$ \citep{Breton2013+4MSPs}
which absorbs several unknown quantities such as the stellar albedo, and the
``beaming factor'' accounting for the pulsar's non-isotropic emission.
A location on the stellar surface which is a distance $r$ from the
pulsar, and whose normal vector is at an angle $\chi$ from the vector pointing
to the pulsar, receives heating power of $\sigma T_{\rm irr}^4 \cos \chi A^2/
r^2$ per unit area. We assume that the star remains in thermal
equilibrium, and so this flux is entirely re-radiated, and hence the surface
temperature at this location is raised to $T = (T_{\rm g}^4 + \cos \chi T_{\rm
  irr}^4 A^2 / r^2)^{1/4}$. To account for the light curve asymmetry and
variability, we require additional parameters describing deviations from this
direct-heating temperature model; these will be discussed below.

Given this set of parameters, \texttt{Icarus} computes model light curves in
each band by solving for the stellar equipotential surface, generating a grid of
elements covering this surface, calculating the temperature of each element as
above, and simulating the projected flux (including limb darkening) from every
surface element at a given inclination angle and at the required orbital
phases. For the flux simulation, we used the model spectra from the
G\"{o}ttingen Spectral
Library\footnote{\url{http://phoenix.astro.physik.uni-goettingen.de/}}
\citep{Husser2013+Atmos} produced by the PHOENIX \citep{Hauschildt+PHOENIX}
stellar atmosphere code. We integrated these model spectra over the transmission
curves of the observing setups to obtain flux models in the ULTRACAM and GROND
filters.

The flux was rescaled in each band for a distance $d$ and reddening due to
interstellar extinction, parameterised by the V-band extinction, $A_{\rm V}$,
for which we assumed a uniform prior between $0.0 < A_{\rm V} < 0.14$, with the
(conservative) upper limit being twice that found by \citet{Romani2015+J2039}
from fits to the X-ray spectrum.  Since the \textit{Gaia} parallax measurement
is marginal, we followed the recommendations of \citet{Luri2018+GAIApi} to
derive a probability distribution for the distance by multiplying the Gaussian
likelihood of the parallax measurement, $p(\varpi | d)$, by an astrophysically
motivated distance prior for MSPs. For this, we take the density of the Galactic
MSP population along the line of sight to J2039 according to the model of
\citet{Levin2013+GalMSPs}. This model has a Gaussian profile in radial distance
from the Galactic centre ($r$) with width $r_0 = 4.5$\,kpc, and an exponential
decay with height $z$ above the Galactic plane, with scale height $z_0 =
0.5$\,kpc. The transverse velocity distribution for binary MSPs in the ATNF
Pulsar Catalogue \citep{Manchester2005+ATNF} is well approximated by an
exponential distribution with mean $v_0 = 100$\,km~s$^{-1}$, which we apply as
an additional distance prior.  In total, the distance prior is,
\begin{equation}
  p(d) \propto p(\varpi \,|\, d)\, d^2\, \exp\left[-\frac{v(d)}{v_0} -
    \frac{z(d)}{z_0} - \frac{1}{2}\left(\frac{r(d)}{r_0}\right)^2\right]\;,
\end{equation}
where the $d^2$ term arises from integrating the Galactic MSP density model at
each distance over the 2D area defined by the \textit{Gaia} localization
region. Finally, we used the radio dispersion measure, $\textrm{DM} =
24.6$\,pc~cm$^{-3}$ (see Paper II) as an additional distance constraint. The
Galactic electron density model of \citet[][hereafter \citetalias{YMW16}]{YMW16}
gives an estimated distance of $d = 1.7$\,kpc, with nominal fractional
uncertainties of $\pm45$\%. We therefore multiplied the distance prior by a
log-normal distribution with this mean value and width. This overall prior gives
a 95\% confidence interval of $1.2\,\textrm{kpc} < d < 3.0\,\textrm{kpc} $, with
expectation value $\hat{d} = 1.9$\,kpc.

In our preliminary \texttt{Icarus} models, constructed prior to the
spectroscopic observations by \citet{Strader2019+RBSpec} and the pulsation
detection presented here, we jointly fit all three light curves, and
additionally fit for $P_{\rm orb}$ and $T_{\rm asc}$. For this we used a
Gaussian prior on $P_{\rm orb}$ according to the best-fitting period and
uncertainty from the Catalina Surveys Southern periodic variable star catalogue
\citep[][see Section~\ref{s:archival}]{Drake2017+CSSPVC}, and refolded the
optical observations appropriately. The resulting posterior distributions on
$P_{\rm orb}$, $T_{\rm asc}$ and on $x$ were used to constrain the parameter
space for the gamma-ray pulsation search in Section~\ref{s:pulsar}.

In these preliminary models, we accounted for the light curve asymmetry and
variability by describing the surface temperature of the star using an empirical
spherical harmonic decomposition whose coefficients could vary between the three
epochs. While this model served our initial goal of phase-aligning the light
curves to constrain the orbital parameters, the spherical harmonic temperature
parameterisation suffered from several deficiencies. Firstly, the decomposition
had to include at least the quadrupole ($l=2$) order to obtain a satisfactory
fit. Several of these coefficients were highly correlated with one another, and
polar terms $(m=0)$ are poorly constrained as the system is only viewed from one
inclination angle, leading to very poor sampling efficiency. Secondly, the
quadrupole term naturally adds power into the second harmonic of the light
curve, changing the amplitude of the two peaks in the light curve. In the base
model, this amplitude depends only on the inclination and Roche-lobe filling
factor, and so the extra contribution of the quadrupole term made these
parameters highly uncertain.

To try to obtain more realistic parameter estimates, we instead modelled the
asymmetry and variability by adding a cold spot to the surface temperature of
the star. While cool star spots caused by magnetic activity are a plausible
explanation for variability and asymmetry in the optical light curves
\citep{vanStaden2016+SpottyRB}, other mechanisms such as asymmetric heating from
the pulsar \citep{Romani2016+IBS,Sanchez2017+Bduct}, or heat re-distribution due
to convective flows on the stellar surface \citep{Kandel2020+ConvFlow, Voisin2020+HeatDist}, may also explain
this. Our choice to model the light curves using a cool spot came from this
being a convenient parameterisation for a temperature variation on the surface
of the star, rather than from assuming that variability is due to magnetic star
spot activity.

In our model, this spot subtracts from the gravity-darkened temperature of the
star, with a temperature difference of $\tau$ at the centre of the spot, which
falls off with a 2D Gaussian profile with width parameter $\rho$ in angular
distance ($\Delta$) from the centre of the spot. The spot location on the
surface of the star is parameterised by the polar coordinates $(\theta,\phi)$,
with $\theta=0$ aligned with the orbital angular momentum, $\phi=0$ pointing
towards the pulsar and $\phi=90\degr$ aligned with the companion's direction of
motion. We assumed a sinusoidal prior on $\theta$ to ensure our priors covered
the surface of the star approximately uniformly (the approximation would be
exact for a spherical, i.e. non-rotating and non-tidally distorted star). The
spot width was confined to be $5\degr < \rho < 30\degr$. The lower limit
prevents very small and very cold spots, while the upper limit ensures that the
effects of spots do not extend over much more than one hemisphere.

To prevent over-fitting, we added an extra penalty factor on the total
(bolometric) difference in flux that the spot adds to the model. This is,
approximately, proportional to $I = \iint_{S}\,\tau^4
e^{-(\Delta(\theta,\phi)^2/2\rho^2)}\,dS$ where $S$ is the surface of the
star. In our fits we adopted a Gaussian prior on $I$, centred on $I=0$ with
width parameter $\sigma_{I} = 6.25\times10^{10}\,\textrm{K}^4\,\textrm{sr}$,
corresponding to a $-500$K spot covering 1 steradian of the star's
surface. Noting from Figure~\ref{f:LC} that the first peak (at the pulsar's
ascending node) is always larger than the second, and that the variability seems
to be strongest around the second peak, we assumed in our model that the light
curve asymmetry is due to a variable cold spot ($\tau < 0$\,K) on the leading
edge of the companion star, and confined $0\degr < \phi < 180\degr$.

To investigate the light curve variability and understand what effect this has
on our inference of the binary parameters, we chose to fit each light curve
separately. Here we only model the three complete light curves from 2014, 2017
and 2018. The partial 2019 light curve is missing the first peak, and hence
models fit only to the data around the variable second peak would have very high
uncertainties on the fit parameters, making this of limited use compared to the
other three light curves. We included the ULTRACAM $u_s$ data in our
  model fitting, since they were obtained simultaneously with the $i_s$ and $g_s$
  data without requiring additional observing time, and provide an additional
  colour for temperature estimation. However, as the signal-to-noise is much
  lower in this band, we do not expect it to have had a large effect on the
  results.

To account for uncertainties in our atmosphere models, extinction, or
photometric calibration, we allowed for constant offsets in the magnitudes in
each band, penalising the chi-squared log-likelihoods using a Gaussian prior on
the magnitude offset with a width of $0.05$\,mag. As the resulting reduced
chi-squared was greater than unity, we also applied rescaling factors to the
uncertainties in each band. Both the band calibration offsets and uncertainty
rescaling factors were computed to maximise the penalised log-likelihood.

At each sampled location in the parameter space, \texttt{Icarus} additionally
computed the projected velocity of every surface element, and averaged these
weighting by their $r^{\prime}$ flux, to obtain a simulated radial velocity
curve. This filter band was chosen as it covers the sodium absorption line seen
in \citet{Strader2019+RBSpec}. The simulated radial velocity curve was compared
to the measured radial velocities from \citet{Strader2019+RBSpec}, additionally
fitting for a constant systemic radial velocity, and the resulting chi-squared
term added to the overall log-likelihood.

\begin{figure*}
	\includegraphics[width=0.95\textwidth]{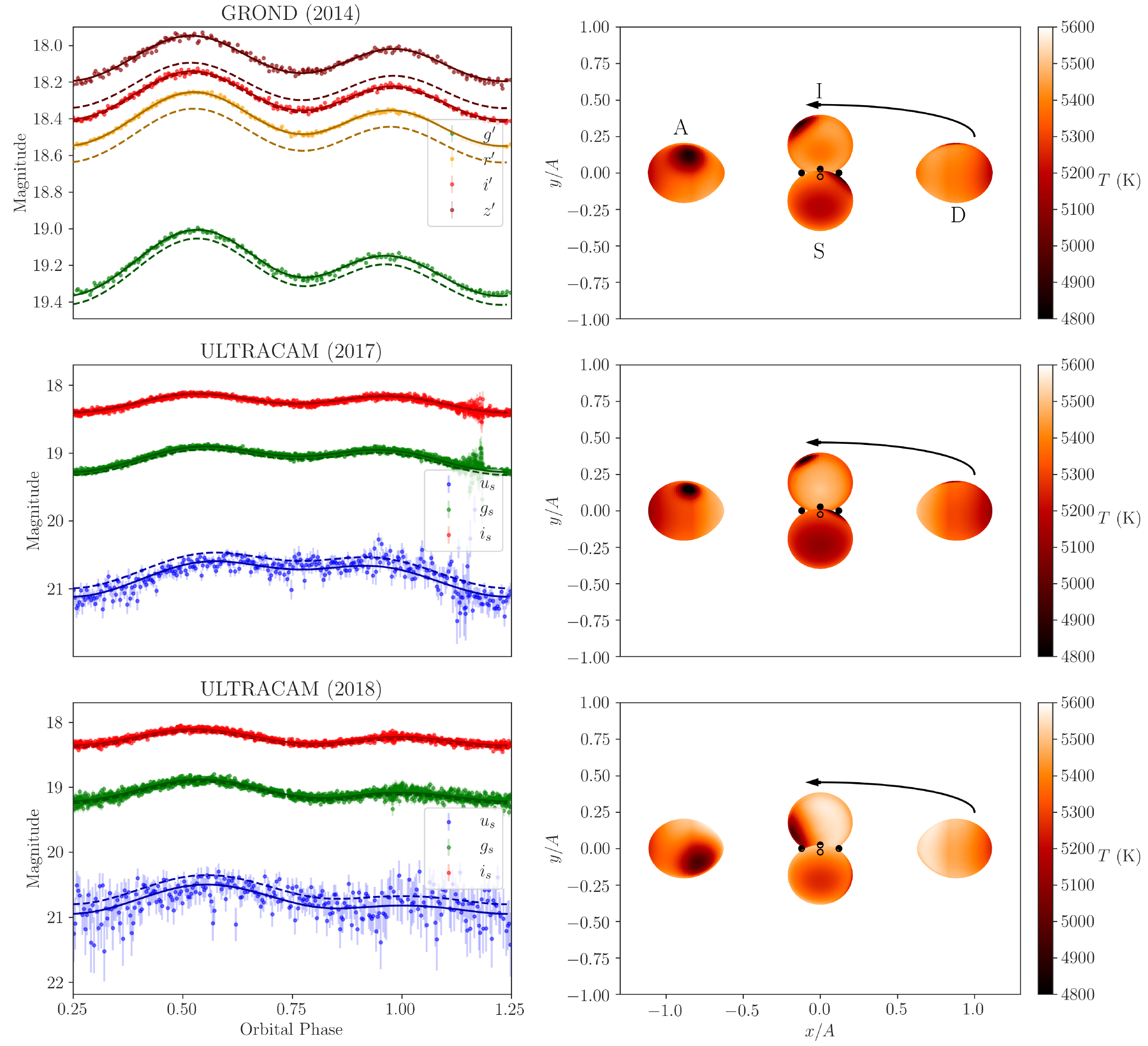}
  \caption{Best-fitting optical light curve models for J2039. Each row shows the
    best-fitting model for a given epoch. Left panels show the observed light
    curves and uncertainties (coloured error bars) in each band. The fluxes in
    each band predicted by the best-fitting \texttt{Icarus} model are shown as
    dashed curves. When fitting these models we allowed for small offsets in the
    flux calibration of the observed light curves. The solid curves show the
    model light curves after applying these calibration offsets. Right panels
    show the \texttt{Icarus} model according to the best-fitting parameters at
    the pulsar's ascending (A) and descending (D) nodes, and superior (S) and
    inferior (I) conjunctions, marked on the top right panel, with the direction
    of motion shown by an arrow. The pulsar's position at each phase is shown by
    a black dot. Phase zero in the light curves corresponds to the pulsar's
    ascending node. The axes are in units of the orbital separation ($A$). The
    surface temperature of the companion star is shown by the colour bar.}
    \label{f:Multinest_model}
\end{figure*}

\begin{figure*}
	\includegraphics[width=0.95\textwidth]{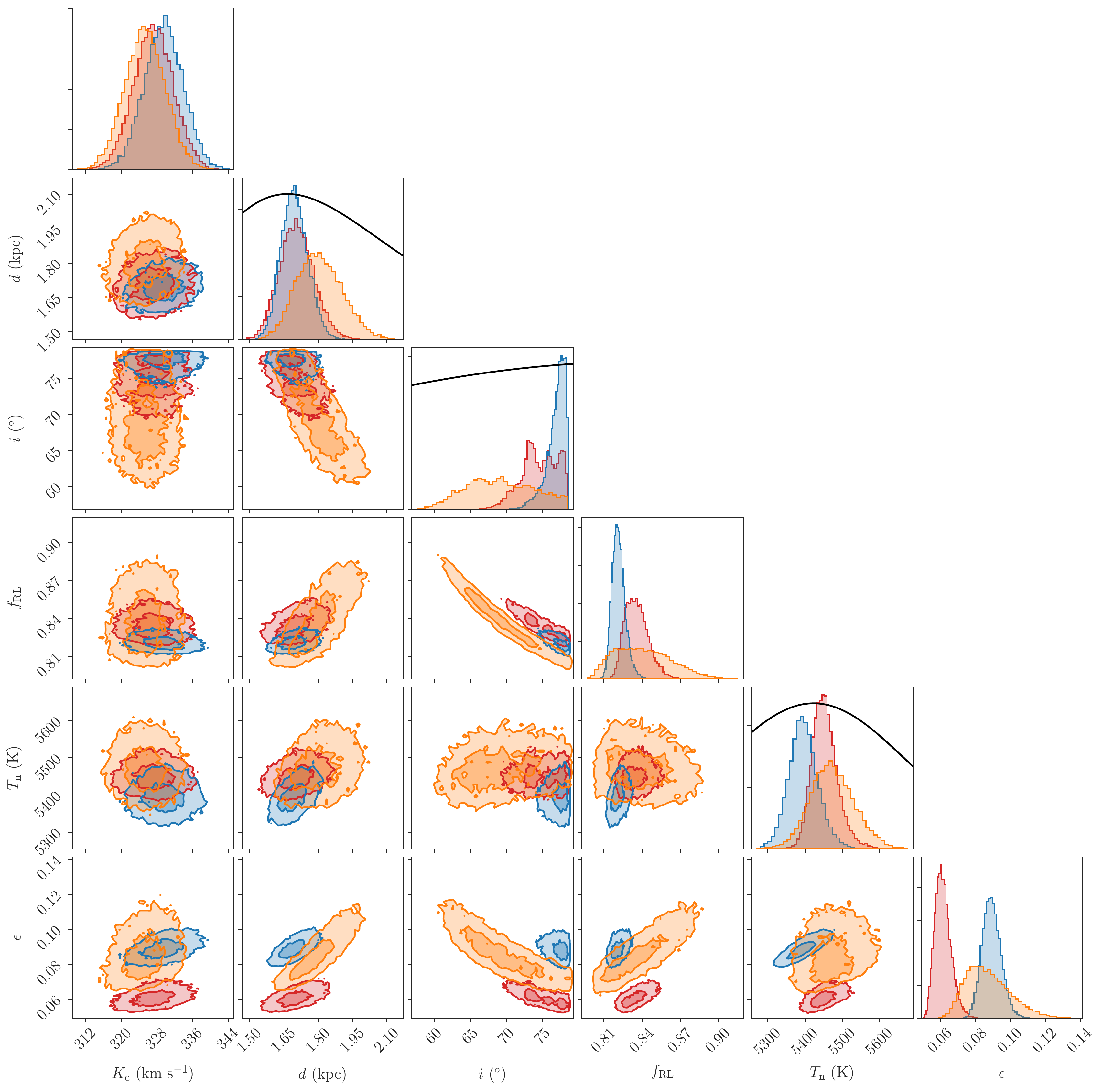}
  \caption{Posterior distributions for the \texttt{Icarus} model parameters. Red, blue and orange histograms and contours show the posterior distributions from fits to the GROND, ULTRACAM (2017) and ULTRACAM (2018) light curves, respectively. Contour lines are shown at $1\sigma$ and $2\sigma$ levels. Where a non-uniform prior is assumed, this is shown as a black curve on the corresponding parameter's 1-dimensional histogram. }
    \label{f:Multinest_params}
\end{figure*}

The model fits were performed using the \texttt{pymultinest} Python interface
\citep{PyMultiNest} to the \texttt{Multinest} nested sampling algorithm
\citep{MultiNest}.
The best fitting models and light curves are shown in
Figure~\ref{f:Multinest_model}, the posterior distributions for our model
parameters are shown in Figures~\ref{f:Multinest_params}, \ref{f:Multinest_mass}
and \ref{f:Multinest_cold_spot}, and numerical results are given in
Table~\ref{t:icarus_res}.  While the inferred posterior distributions
  from each epoch generally overlap with each other (except for the spot and
  heating parameters encapsulating variability), in the following discussion we
  take the full range covered by the 95\% confidence intervals of the three
  posterior distributions as estimates for the model uncertainty, in the hope
  that biases due to variability are contained within that range.

\begin{table*}
    \centering
    \caption{\texttt{Icarus} fit results. Numerical values are the median of the marginalised posterior distributions output by \texttt{Multinest}, with the $95$\% confidence regions shown in sub- and superscript. }
    \label{t:icarus_res}
    {\renewcommand{\arraystretch}{1.2}
    \begin{tabular}{lccc}
      \hline
      Parameter & 2014 June 16--18 (GROND) & 2017 June 18 (ULTRACAM) & 2018 June 02 (ULTRACAM)\\
      \hline
 $\chi^2$ (degrees of freedom) & $958.6\,(824)$ & $3930.3\,(3529)$ & $3699.5\,(3330)$\\
      \hline
      \multicolumn{4}{c}{\texttt{Icarus} fit parameters}\\
      \hline
Systemic velocity (km s$^{-1}$) & $6_{-6.5}^{+6.6}$ & $5.9_{-6.4}^{+6.4}$ & $6.4_{-6.6}^{+6.7}$ \\
Companion's projected radial velocity, $K_{\rm c}$ (km s$^{-1}$) & $327.3_{-8.8}^{+8.9}$ & $329.6_{-8.7}^{+8.6}$ & $325.4_{-8.9}^{+9}$ \\
Distance, $d$ (kpc) & $1.7_{-0.14}^{+0.16}$ & $1.69_{-0.11}^{+0.12}$ & $1.8_{-0.17}^{+0.21}$ \\
V-band extinction, $A_V$ & $0.072_{-0.069}^{+0.064}$ & $0.091_{-0.082}^{+0.047}$ & $0.095_{-0.086}^{+0.043}$ \\
Inclination, $i$ ($\degr$) & $74.5_{-5.1}^{+3.7}$ & $77.1_{-3.8}^{+1.2}$ & $69_{-8}^{+8.8}$ \\
Roche-lobe filling factor, $f_{\rm RL}$ & $0.835_{-0.014}^{+0.021}$ & $0.821_{-0.008}^{+0.012}$ & $0.839_{-0.033}^{+0.048}$ \\
Base temperature, $T_{\rm n}$ (K) & $5451_{-57}^{+82}$ & $5395_{-74}^{+87}$ & $5471_{-109}^{+124}$ \\
Irradiating temperature, $T_{\rm irr}$ (K) & $3456_{-72}^{+96}$ & $3807_{-85}^{+94}$ & $3700_{-141}^{+160}$ \\
Spot central temperature difference, $\tau$ (K) & $-540_{-150}^{+170}$ & $-620_{-260}^{+340}$ & $-600_{-80}^{+70}$ \\
Spot Gaussian width parameter, $\rho$ ($\degr$) & $23_{-3.9}^{+5.7}$ & $14.4_{-3.7}^{+10.9}$ & $28_{-3.9}^{+1.9}$ \\
Spot co-latitude, $\theta$ ($\degr$) & $42.7_{-7.5}^{+9.1}$ & $33.1_{-10.5}^{+16.1}$ & $96.7_{-13.1}^{+13.2}$ \\
Spot longitude, $\phi$ ($\degr$) & $77.9_{-4.6}^{+3}$ & $73.6_{-12.4}^{+6.2}$ & $56.3_{-4.6}^{+3.8}$ \\
      \hline
      \multicolumn{4}{c}{Derived parameters}\\
\hline
Heating efficiency, $\epsilon$ & $0.061_{-0.007}^{+0.011}$ & $0.089_{-0.01}^{+0.012}$ & $0.086_{-0.019}^{+0.03}$ \\
Pulsar mass, $M_{\rm psr}$ ($M_{\odot}$) & $1.2_{-0.11}^{+0.14}$ & $1.18_{-0.09}^{+0.1}$ & $1.3_{-0.2}^{+0.31}$ \\
Companion mass, $M_{\rm c}$ ($M_{\odot}$) & $0.165_{-0.012}^{+0.017}$ & $0.162_{-0.008}^{+0.011}$ & $0.18_{-0.026}^{+0.041}$ \\
Mass ratio, $q \equiv M_{\rm psr}/M_{\rm c}$ & $7.27_{-0.2}^{+0.2}$ & $7.32_{-0.19}^{+0.19}$ & $7.22_{-0.2}^{+0.2}$ \\
Volume-averaged companion density (g cm$^{-3}$) & $4.38_{-0.14}^{+0.1}$ & $4.48_{-0.09}^{+0.07}$ & $4.35_{-0.27}^{+0.25}$ \\
Spot integral, $I$ ($10^{10}$ K$^4$ sr) & $8.4_{-5.7}^{+8.8}$ & $6.0_{-5.3}^{+9.4}$ & $18.1_{-6.5}^{+8.3}$ \\
\hline
    \end{tabular}
    }
\end{table*}

\section{Results and Discussion}
\label{s:discussion}
\subsection{Binary Inclination and Component Masses}
\begin{figure}
	\includegraphics[width=\columnwidth]{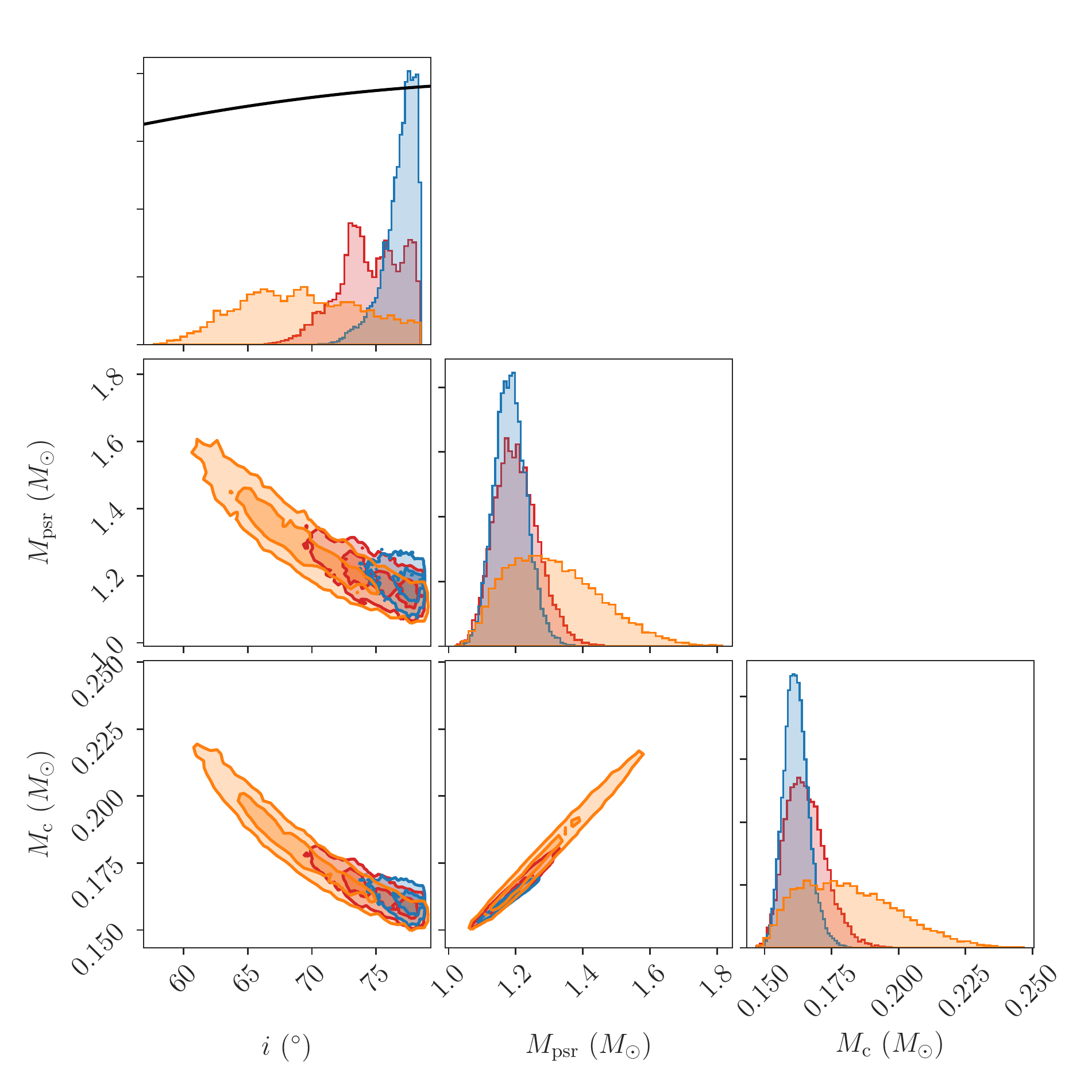}
  \caption{Posterior distributions from \texttt{Icarus} model fitting, as in
    Figure~\ref{f:Multinest_params}, but for the binary inclination, pulsar and
    companion masses.}
    \label{f:Multinest_mass}
\end{figure}

Perhaps one of the more important questions is whether or not we are able to
obtain a reliable measurement of the mass of the neutron star in the system. The
maximum neutron star mass is a crucial unknown quantity which can discriminate
between different nuclear equations-of-state \citep[see ][and references
  therein]{Ozel2016+EoS}. Recent works \citep{vanKerkwijk+B1957,Linares2018+J2215} have
found very heavy pulsar masses for spider pulsars, and there are hints that
these systems may in general contain heavier neutron stars than e.g. double
neutron star systems \citep{Strader2019+RBSpec}.

Combining the radial velocity curve measured via spectroscopy by
\citet{Strader2019+RBSpec} with our pulsar timing measurement of the pulsar's
projected semi-major axis constrains the mass ratio to $q = 7.3 \pm 0.2$. Hence, all parameters in the binary mass function
(Equation~\ref{e:mass_function}) are relatively well measured, with the
exception of the binary inclination angle. Measuring this by modelling the
optical light curves was therefore a key goal for our study of this system. The
observed asymmetry and variability in the light curve are significant
complicating factors for this, as estimates for the inclination angle are
determined by the amplitude of the ellipsoidal peaks. In J2039, these do not
have equal amplitudes, and vary over time.

Our optical model fits to all three complete orbital light curves consistently
preferred high inclinations, hitting the upper limit of ($i \lesssim 79 \degr$)
imposed by our assertion that the pulsar is not eclipsed at superior
conjunction. The second ULTRACAM light curve results in the widest $95$\%
confidence interval, with $61 \degr < i < 78\degr$. Marginalising over the
uncertainty in the radial velocity amplitude, the corresponding pulsar mass
range is $1.1 M_{\odot} < M_{\rm psr} < 1.6 M_{\odot}$, with a median of $M_{\rm
  psr} \approx 1.3 M_{\odot}$, but the models for the other two epochs give
narrower ranges $1.1\, M_{\odot} < M_{\rm psr} < 1.35\, M_{\odot}$. The
posterior distributions on these parameters are shown in
Figure~\ref{f:Multinest_mass}.

Inclination angles derived from optical light curve fits are highly dependent on
the chosen temperature and irradiation models and priors. In particular, we
caution that there is likely to be a large (but unknown) systematic uncertainty
underlying our inclination estimates, caused by our simplifying assumption that
the variability and asymmetry can be modelled by one cold spot on the leading
face of the star. Other models for light curve asymmetry, e.g. intra-binary
shock heating models or models featuring convective winds on the stellar surface
\citep{Romani2016+IBS,Kandel2020+ConvFlow,Voisin2020+HeatDist}, may give
different results. Pulsar masses derived from optical light curve modelling
should therefore be treated with caution, as the results can be highly
model-dependent.  For instance, if part of the asymmetry is caused by excess
  heating on the trailing face of the star rather than a cool spot on
    the leading face, then the leading peak of the light curve will be larger
  than predicted by the direct-heating model, and the model's inclination angle
  will increase to compensate. Nevertheless, our results suggest that a high
  inclination and fairly low pulsar mass is compatible with the
  observed light curves.

Our resulting mass is rather
  lower than those inferred from other redback systems, which
  \citet{Strader2019+RBSpec} found to cluster around $1.8\,M_{\odot}$,
but has a range similar to that found for PSR~J1723$-$2837
($M_{\rm psr} < 1.4\,M_{\odot}$) by \citet{vanStaden2016+SpottyRB}. While some
of the redback masses compiled by \citet{Strader2019+RBSpec} do have strict
lower limits (i.e. for edge-on orbits) that are above our inferred mass range,
it is possible that unmodelled asymmetries and variability may be systematically
biasing optical-modelling based inclination measurements to lower values, and
hence biasing the redback pulsar mass distribution towards higher values.

By generating and fitting a flux-averaged radial velocity curve, our binary
system model additionally corrects for possible biases in the observed radial
velocity curve due to a difference between the centre of mass of the companion
star and the position on the surface where spectral lines contribute most
strongly to the observed spectra \citep[e.g.,][]{Linares2018+J2215}. For J2039,
heating has a fairly small effect on the light curve, and the resulting
correction to the radial velocity curve is small: the epoch with the largest
inferred centre-of-mass radial velocity amplitude (2017 June 18) has $K_2 = 330
\pm 5$\,km\,s$^{-1}$, compared to $K_2 = 324 \pm 5$\,km\,s$^{-1}$ that
\citet{Strader2019+RBSpec} found from a simple sinusoidal fit. This implies that
the required $K_2$-correction is only $\Delta K_2/K_2 \lesssim 2$\%, and
only increases the
  inferred pulsar mass by $\Delta M_{\rm psr} / M_{\rm psr} \lesssim
6$\%. While here this additional bias is far lower than that caused by our
uncertainty on the inclination, this is not true in general for other redback
systems. Large changes in the heating of redback companions have been observed
\citep{Cho2018+VariableRBs}, and so reliable centre-of-light corrections require
photometry observations to be taken as close in time as possible to
spectroscopic radial velocity measurements to mitigate possible errors due to
variations in heating.

Our pulsar mass range is lower than that estimated by \citet{Strader2019+RBSpec}
($M_{\rm psr} > 1.8 M_{\odot}$) from similar fits to the GROND light
curve. Prior to our pulsation detection, the binary mass ratio was
unconstrained, and so this was an additional free parameter in their model. The
authors used two large cold spots in their model, which were both found to lie
towards the unheated side of the star. These spots will affect the amplitudes of
both ellipsoidal peaks, and therefore will affect the estimation of the
inclination angle, filling factor and mass ratio that are constrained by these
amplitudes. Their fits found a much lower mass ratio than is obtained from the
pulsar's semi-major axis measurement ($q < 5.3$ vs. $q = 7.3 \pm 0.2$ here) and
a nearly Roche-lobe filling companion $f_{\rm RL} \approx 95$\%. Both of these
parameter differences will increase the amplitude of the ellipsoidal
modulations, allowing for a more face-on inclination and thus a heavier pulsar,
explaining our disagreement.

The inferred inclination angle is also (qualitatively) consistent with the
observed gamma-ray pulse profile. Since the pulsar has been spun-up via
accretion its spin axis should be aligned to the orbital axis, and hence the
pulsar viewing angle (the angle between the line-of-sight and the pulsar's spin
axis) will match the orbital inclination. The gamma-ray pulse profile features
one broad main peak, with a smaller trailing peak. This therefore rules out an
equatorial viewing angle to the pulsar, and hence an edge-on orbital inclination
$i\sim90\degr$ as in that case the gamma-ray pulse should exhibit two similar
peaks approximately half a rotation apart. The detection of radio pulsations
enables a full investigation of this, fitting both the gamma-ray pulse profile
shape and its phase relative to the radio pulse using theoretical pulse emission
models. This will be described in detail in Paper II, but we note here that
these models suggest a lower viewing angle of $i \sim 67\degr$, for a
pulsar mass of $M_{\rm psr} \sim
1.4\,M_{\odot}$.

For the companion mass, we find $0.15 M_{\odot} < M_{\rm c} < 0.22 {\rm
  M}_{\odot}$. Our \texttt{Icarus} model fits gave the companion star base
temperature $T_{\rm n} \approx 5400\,$K and volume-averaged radius $R_{\rm c}
\approx 0.4$\,R$_{\odot}$. These are both significantly larger than would be
expected for a main-sequence star of the same mass. Indeed, this is not
surprising, as the accretion required to recycle the MSP will have stripped the
majority of the stellar envelope, while tidal forces and heating from the pulsar
continue to add additional energy into the companion star
\citep{Applegate1994+B1957}, causing a further departure from ordinary stellar
evolution.

\subsection{Distance and Energetics}
The \texttt{Icarus} fits to our three light curves all returned consistent
distance estimates around $d = 1.7^{+0.3}_{-0.1}$~kpc, consistent with the \textit{Gaia}
parallax and \citetalias{YMW16} DM distance estimates.  Assuming a fiducial
distance of $d = 1.7$~kpc, the \textit{Gaia} proper motion measurement implies a
transverse velocity of $v_{\rm T} \approx 125$\,km s$^{-1}$. This transverse
velocity will induce an apparent linear decrease in both the spin and orbital
frequencies due to the increasing radial component of the initially transverse
velocity \citep[hereafter referred to as the Shklovskii effect
  after][]{Shklovskii}. This effect accounts for around $20$\% of the observed
spin-down rate. An additional contribution to the observed spin-down rate comes
from the pulsar's relative acceleration due to the Galactic rotation and
gravitational potential. Using the formula given by
\citet{Matthews2016+Nanograv} and references therein, we estimate this accounts
for less than 1\% of the observed spin-down rate. At the fiducial distance the
gamma-ray flux corresponds to a luminosity of $L_{\gamma} =
5\times10^{33}$\,erg\,s$^{-1}$, or a Shklovskii-corrected gamma-ray efficiency
of $\eta_{\gamma} = L_{\gamma}/\dot{E} = 21$\%, which is typical for gamma-ray
MSPs \citep{2PC}. Recently, \citet{Kalapotharakos2019+FP} discovered a
``fundamental plane'' linking pulsars' gamma-ray luminosities to their spin-down
powers, magnetic field strengths and spectral cut-off energies
\citep{Kalapotharakos2019+FP}. For J2039, this predicts $L_{\gamma,\textrm{FP}}
= 1.3\times10^{34}\,\textrm{erg s}^{-1}$, or $0.4\,\textrm{dex}$ above the
observed value, consistent with the scatter about the fundamental plane seen by
\citet{Kalapotharakos2019+FP}.

In our \texttt{Icarus} model, we assume that the inner side of the companion
star is heated directly by flux from the pulsar. For PSR~J2039$-$5617, our
optical models hint that the heating flux reaching the companion star may be
variable, and is on the order of a few percent of the total spin-down luminosity
of the pulsar, with $\epsilon = L_{\rm irr}/\dot{E} \sim 0.05$ to $0.12$. This is a somewhat lower efficiency than is typically observed in
spider systems, where heating normally accounts for around $20$\% of the
pulsar's spin-down power \citep{Breton2013+4MSPs,Draghis2019+BWs}.

The precise nature of the mechanism by which redback and black-widow pulsars
heat their companions is currently unclear. For J2039, the inferred gamma-ray
luminosity is larger than the heating power, and so we may infer that gamma rays
are a sufficient heating mechanism in this case. For other spiders, this is not
always true, with heating powers found to be much larger than gamma-ray
luminosities \citep[e.g.,][]{Nieder2019+J0952}. Some discrepancy between the two
can be explained by underestimated distances, or beamed (i.e. non-isotropic)
gamma-ray flux that is preferentially emitted in the equatorial plane, although
heating efficiencies and gamma-ray efficiencies remain only loosely correlated
even with these corrections \citep{Draghis2019+BWs}. This may indicate that
another mechanism, e.g. high-energy leptons in the pulsar wind, is responsible
for heating the companion star. Note that both $\eta_{\gamma}$
and $\epsilon$ are fractions of $\dot{E}$, so while $\dot{E}$ is an
order-of-magnitude estimate dependent on the chosen value for the pulsar moment
of inertia, the ratio between $\eta_{\gamma}$ and $\epsilon$ is independent of
this. 

\subsection{Optical light curve asymmetry and variability}
In the above heating efficiency calculation, we only included \textit{direct}
heating i.e. flux from the pulsar that is immediately thermalised and
re-radiated from the surface of the companion star at the location on which it
impinges. For J2039 the asymmetry of the light curve, and relative lack of
variability on the leading peak may suggest that some heating is being
re-directed toward the trailing face of the companion star, keeping this side at
a more constant temperature. However, with only three optical light curves
covering this orbital phase this is purely speculative, and requires additional
optical monitoring to check for variability in the leading peak.

Nevertheless, similar light curve asymmetry, with the leading peak typically
appearing as the brighter of the two, seems to be common in many types of close
binary systems (e.g. cataclysmic variables (CVs) and W UMa-type eclipsing
binaries), where it is often referred to as the \textit{O'Connell effect}
\citep[after][]{OConnellEffect}. Several processes have been proposed to explain
this in general, and in redbacks in particular, but so far without
consensus. Possible processes include: reprocessing of the pulsar wind by a
swept-back asymmetric intra-binary shock \citep{Romani2016+IBS}; channeling of
charged particles in the pulsar wind onto the poles of a companion's misaligned
dipolar magnetic field \citep{Sanchez2017+Bduct,Wadiasingh2018}; or heat
redistribution due to fluid motion in the outer layers of the star
\citep{Martin1995+convection,Kandel2020+ConvFlow,Voisin2020+HeatDist}.

For J2039, the presence of an intra-binary shock wrapping around the pulsar is
required to explain the observed orbital modulation of X-rays. Following the
model of \citet{Romani2016+IBS}, it therefore seems plausible that extra heating
flux could be directed at the trailing face of the companion star, and could at
least partially explain the observed light-curve asymmetry. We are then left to
explain the variability in the light curve. \citet{Cho2018+VariableRBs} observe
similar variability in the light curves of three other redback systems,
attributing this to variability in the stellar wind and hence in the
intra-binary shock.

An alternative explanation for redback variability is that magnetic activity in
the companion leads to large cool star spots on the stellar surface, which
migrate around the star and may appear and disappear over time. This star-spot
interpretation has been invoked to explain the similar optical variability seen
in long-term monitoring of the redback system PSR~J1723$-$2837
\citep{vanStaden2016+SpottyRB}. A periodogram analysis of these light curves
found a component with a period slightly shorter than the known orbital period,
which the authors interpret as being due to asynchronous (i.e. non-tidally
locked) rotation of the companion star. Alternatively, this could also be due to
differential rotation of the stellar surface, as seen in sun spots, and observed
e.g. in CV secondaries via Roche tomography
\citep[e.g.,][]{Hill2014+CVDiffRot}. Given the year-long time intervals between
our ULTRACAM light curves of J2039 we cannot perform the same analysis to track
a single variable component over time to confirm this picture, but this may be
possible in the future with sufficient monitoring. Another interesting question
that may be addressed with additional monitoring is whether or not the optical
variability correlates with the variations in the orbital period, as both may be
linked through magnetic cycles in the stellar interior.

\begin{figure}
	\includegraphics[width=\columnwidth]{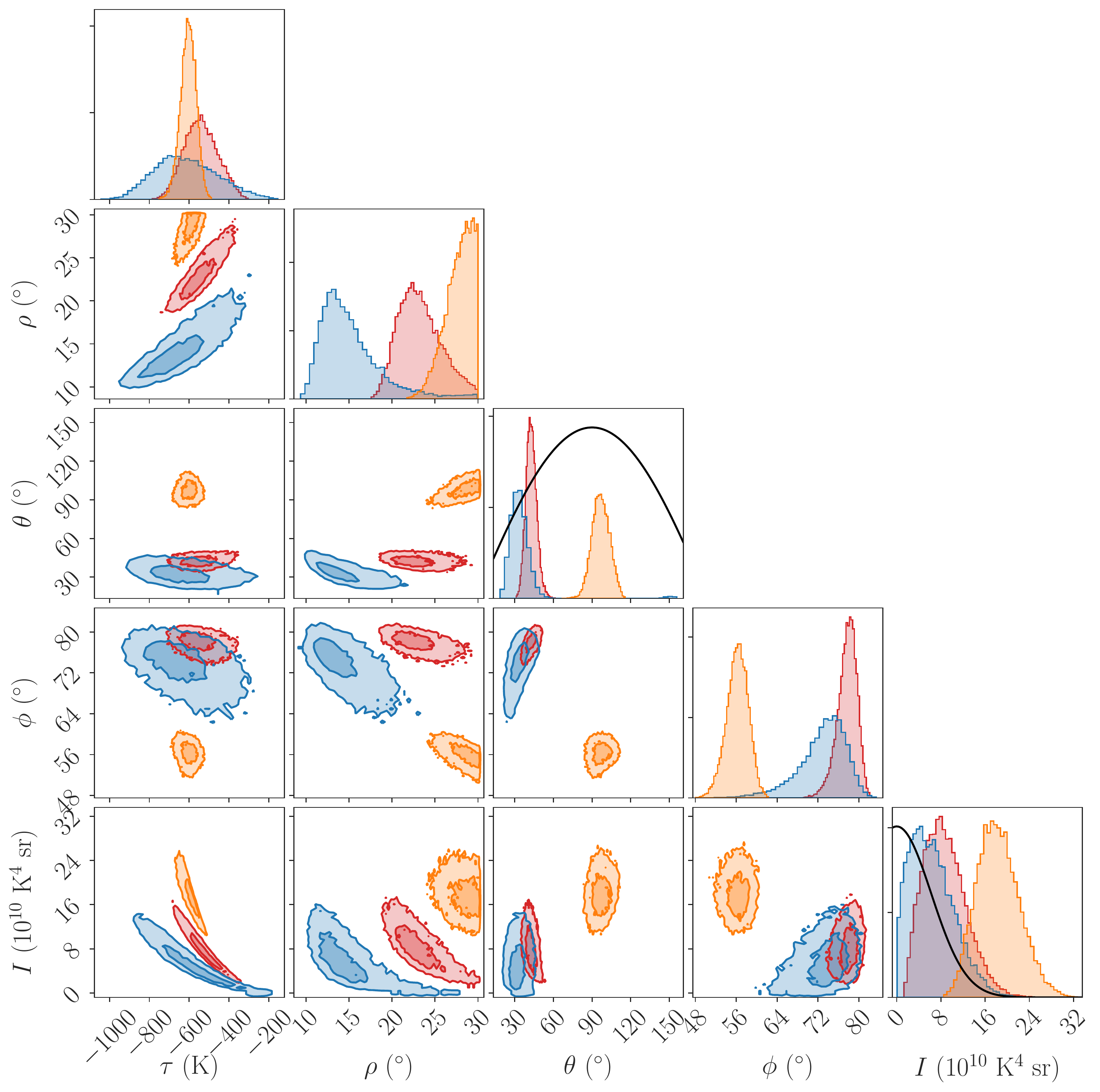}
  \caption{Posterior distributions from \texttt{Icarus} model fitting, as in
    Figure~\ref{f:Multinest_params}, but for the parameters of the cold spot
    added to the companion's surface. These parameters account for the
    significant variability observed in the light curves, hence the rather
    different values recovered.}
    \label{f:Multinest_cold_spot}
\end{figure}

To create our binary system models, we used a toy model for the stellar surface
temperature that included a variable cold spot to account for the asymmetry and
variability. The posterior distributions on the parameters of these spots are
shown in Figure~\ref{f:Multinest_cold_spot}. This model is certainly an
over-simplification of the truth, and so we will avoid placing much emphasis on
the numerical results for these parameters, noting that our goal was instead to
marginalise over the variability to retrieve estimates for more tangible
quantities such as the inclination and filling factor. Our chosen prior, which
aims to minimise the bolometric flux $\propto \tau^4 \sigma^2$ subtracted by the
cool spot, penalises small but very cold spots over larger and warmer
spots. This prevents our model reaching the very cold spot temperatures ($\tau
\sim -2000$\,K) that have been observed in well-studied main-sequence stars
\citep{StarspotReview}. Instead, our model prefers large spots (close to our
upper limit of $\rho = 30\degr$) with a central temperature difference between
$\tau \sim -300$\,K to $\tau\sim -700$\,K. While such a temperature reduction
could be plausibly explained by magnetic star spot activity, we are hesitant to
interpret these as ``true'' star spots, but rather consider them to be areas of
decreased temperature due to unknown variable effects, e.g. asymmetric heating
from the pulsar, or heat re-distribution due to convective flows on the stellar
surface. Continued photometric monitoring of J2039 to test the star-spot
explanation may reveal evidence that these cool areas migrate across the surface
of the star, as they do in PSR~J1723$-$2837
\citep{vanStaden2016+SpottyRB}. We discuss this possibility further
  below. Furthermore, a dedicated study of the spectra observed by
\citet{Strader2019+RBSpec} may be able to detect the presence of spectral lines
associated with cooler temperatures to further investigate the star-spot
hypothesis.

We also note that a better understanding of variability in rotationally powered
redback systems may offer insight into some of the most extreme behaviour
exhibited by binary MSP systems: the sudden (dis)appearance of accretion discs
in transitional MSP systems
\citep[tMSPs,][]{Archibald2009+J1023,Papitto2013+M28I,Bassa2014+J1227,Stappers2014+J1023}.
To provide material to power a tMSP's accretion state, the companion star must
be overfilling its Roche lobe. However, optical modelling of PSR~J1023$+$0038
somewhat surprisingly suggests a companion that significantly underfills its
Roche lobe \citep{McConnell2015+J1023,Shahbaz2019+J1023}. This therefore
requires a significant change in the radius of the companion star, and the
timescale on which this takes place is currently unknown. For J2039,
we also find that the companion star is significantly smaller than its Roche
lobe ($f_{\rm RL} \approx 0.83$), and do not find any evidence for variations in
the stellar radius over the three light curves.

\subsection{Orbital Period Variability}
\label{s:opv}
In Section~\ref{s:timing} we measured the orbital period of J2039,
finding significant deviations in the orbital phase from a constant-period
model. Such variations are common among redback systems
\citep[e.g.,][]{Deneva2016+J1048, Archibald2009+J1023, Pletsch2015+J2339}. This
phenomenon has been attributed to the Applegate mechanism
\citep{Applegate1987,Applegate1994+B1957}, originally invoked to
explain period variations in eclipsing Algol-type and CV binaries, in which
periodic magnetic activity cycles in the convective zone of the companion star
introduce a varying quadrupole moment, which couples with the orbital angular
moment to manifest as variations in the orbital period.

Using our new Gaussian process description for the orbital phase variations, we
can hope to quantify the required changes in the quadrupole moment using the
best-fitting values for the hyperparameters of the Gaussian process used to
model the orbital phase variations in Section~\ref{s:timing}.

Under the Applegate model, the change in orbital period is directly related to
the change in the companion star's gravitational quadrupole moment $Q$
\citep{Applegate1987},
\begin{equation}
  \frac{\Delta P_{\rm orb}}{P_{\rm orb}} = -9\frac{\Delta Q}{M_{\rm c} A^2}\,,
\end{equation}
where $A = x (1 + q) / \sin i$ is the orbital separation. For comparison, the
total quadrupole moment induced by the spin of the companion star and the tidal
distortion in the pulsar's gravitational field is \citep{Voisin2020+TimingModel}
\begin{equation}
  \frac{Q}{M_{\rm c} A^2} = -\frac{2}{9} k_2 \left(\frac{R_{\rm c}}{A}\right)^5 \left(4 q + 1\right)\,,
\end{equation}
where $R_c$ is the radius of the companion star and $k_2$ is the apsidal motion
constant, a parameter describing the deformability of the companion star
\citep{Sterne1939+Apsidal}. For solar-type stars $k_2 \sim 0.035$
\citep{Ogilvie+Sunk2}, while if we assume that redback companions are akin to
the companions in CV systems whose outer
envelopes have also been stripped through accretion then we may expect a smaller
value $k_2 \sim 10^{-3}$ \citep{Cisneros-Parra+k2}. For J2039, the
hyperparameter $h = 3.9^{+2.2}_{-1.2}$\,s corresponds to the typical fractional
amplitude for the variations in orbital phase. Taking the simpler squared
exponential covariance function of Equation~(\ref{e:sqexp_cov}) corresponding to
$n \to \infty$ then the deviations in orbital period have covariance function,
\begin{equation}
\begin{split}
  K_{\Delta P_{\rm orb}/P_{\rm orb}}(t_1,t_2) &= \frac{\partial^2 K}{\partial t_1 \partial t_2} \\
  &= \frac{h^2}{l^2} \exp\left(-\frac{(t_1 - t_2)^2}{2\ell^2}\right) \left(1 -
  \frac{(t_1 - t_2)^2}{\ell^4}\right)\,.
\end{split}
\end{equation}
The typical (fractional) amplitude of the orbital period variations is therefore
$\Delta P_{\rm orb}/P_{\rm orb} \sim h / \ell = (3 \pm 1)\times10^{-7}$,
corresponding to $\Delta Q / Q \sim 3\times10^{-5} k_2^{-1}$. The time-varying
component to the gravitational quadrupole moment is therefore required to be of
order a few percent of the total expected quadrupole moment at most to explain
the observed orbital period variations. From this, it seems plausible that the
observed period variations can be powered by quadrupole moment changes, without
requiring that a large fraction of the star be involved in the process. The
required fractional quadrupole moment changes are very similar to those recently
calculated for the companion to the black widow PSR~J2051$-$0827 by
\citet{Voisin2020+J2051}, despite the large difference in their masses.

For our assumed Mat\'{e}rn covariance function, the parameter $n$ is related to
how smooth the noise process is: random walks in orbital phase, period or period
derivative would manifest as noise processes with $n = 1/2,\,3/2$ or $5/2$
respectively \citep{Kerr2015+FermiTiming}. This hyperparameter may therefore
encode information about the source of the orbital period variation. If the
quadrupole moment exhibits random walk behaviour (i.e. the stellar structure
switches rapidly between different states), we would expect to see a random walk
in orbital period ($n=3/2$). Alternatively, if the system is affected by a
variable torque (e.g. variable mass loss, or magnetic braking) then this would
manifest as a random walk in the orbital period derivative or higher orders ($n
\gtrsim 5/2$).

Unfortunately, Figure~\ref{f:opv_psd} illustrates that we are insensitive to the
value of $n$, as the variability quickly falls below the measurement uncertainty
level for periods shorter than $\ell \approx 130$\,d, preventing measurement of the power-law slope above the corner frequency. We find only that a very
shallow power-law spectrum $n < 1$ is ruled out with $95\%$ confidence, but
models with finite $n > 1.5$ fit the data equally well as the squared exponential
kernel corresponding to $n \to \infty$.

We also find marginal evidence for an excess in the noise power at periods
longer than the 11-years of \textit{Fermi}-LAT data. This is not well accounted
for by a longer correlation timescale $\ell$ and shallower spectral index, as
this leaves excess power at intermediate frequencies, and we do find that a
break in the spectrum is preferred, with $\ell \ll T_{\rm obs}$. One explanation
could be that instead of breaking to a constant power level at low frequencies,
the noise process breaks to a shallower power law. However, with only a
handful of independent frequencies below the corner frequency, this slope is
hard to probe, although this may be worth revisiting as the timing
baseline grows.

Alternatively, this low-frequency excess could be explained by a steadily
increasing orbital period, which would introduce a quadratic term in the orbital
phase that would appear as noise power at a period longer than the observation
timespan. In Section~\ref{s:timing}, we accounted for this by
including a constant orbital period derivative $\dot{P}_{\rm orb} = (8 \pm
5)\times10^{-12}$.

While there are several physical processes which could lead to a long term
increase in the orbital period on top of the Applegate-style stochastic
variability, the magnitude of the effect here is hard to reconcile. For example,
the Shklovskii-induced orbital period derivative is $\dot{P}_{\rm orb, Shk} = v_{\rm
  T}^2 P_{\rm orb} / c d = 2\times10^{-14}$, almost three orders of magnitude
smaller than the measured value. Other incompatible explanations for an apparent period
derivative include acceleration in the Galactic potential ($\dot{P}_{\rm orb, acc} = -6 \times 10^{-16}$), or loss of angular momentum due to
gravitational wave emission, which would decrease the orbital period and hence
has the wrong sign here.

In principle, a long-term increase in the orbital period could be explained by
steady mass loss from the system. Under this model, the inferred mass-loss rate
would be $\dot{M} = -0.5\,(M_{\rm c} + M_{\rm psr}) \,\dot{P}_{\rm orb} / P_{\rm
  orb} = -8\times10^{-9}\,M_{\odot}$\,yr$^{-1}$. This is an extremely high
rate, and implies that the companion star would be completely ablated after just
$19$~Myr, assuming a constant mass-loss rate. If we assume that such a mass loss
is driven by material ablated from the companion star by the pulsar, then the
total power budget available for this process should be similar to the spin-down
power of the pulsar. Centrifugal effects from the orbital motion reduce the
gravitational potential difference which must be overcome for matter to escape the system. Denoting the potential at the stellar surface, and the maximum potential within the system as $\varphi_{\rm c}$ and $\varphi_{\rm esc}$, respectively, then an estimate for the maximum possible mass-loss rate due to ablation, assuming 100\% efficiency and isotropic emission from the pulsar, is,
\begin{equation}
  \dot{M}_{\dot{E}} = \frac{\dot{E}\, R_{\rm c}^2}{4 A^2}\frac{1}{\varphi_{\rm c} - \varphi_{\rm esc}}\,.
\end{equation}
Calculating $\varphi_{\rm c}$ and $\varphi_{\rm esc}$ using \texttt{Icarus}, we
find for J2039 $\left|\dot{M}_{\dot{E}}\right| \lesssim
1.5\times10^{-8}\,M_{\odot}$~yr$^{-1}$, and so at first glance it seems that mass-loss
through ablation by the pulsar may be sufficient to explain the observed
$\dot{P}_{\rm orb}$. However, studies of radio eclipses in redback and black widow
systems, in which radio pulsations are absorbed, dispersed and scattered by
diffuse plasma in an extended region outside the companion star's Roche lobe,
typically infer mass-loss rates on the order of $\dot{M} \sim
10^{-12}\,M_{\odot}$~yr$^{-1}$ or lower
\citep[e.g.,][]{Polzin2018+J1810,Polzin2019+J2051}. These mass-loss rates are
therefore clearly incompatible with a mass-loss interpretation for the potential
long-term period increase. Radio eclipses have been observed from J2039, and these
will be investigated in Paper II.

Another alternative mechanism that could lead to a significant $\dot{P}_{\rm orb}$ is that
considered by \citet{vanStaden2016+SpottyRB}, in which asynchronous rotation of the
companion star leads to a tidal force that transfers angular momentum from the
star to the orbit. If the star spins down at a constant rate, $\dot{\Omega}_{\rm
  c}$, then conserving total angular momentum gives
\begin{equation}
  \dot{P}_{\rm orb} = -3 I_{\rm c}\, \dot{\Omega}_{\rm c}\, M_{\rm psr}\, M_{\rm c}
  \left(\frac{2 \pi\, (M_{\rm psr} + M_{\rm c})\, P^2}{G^2}\right)^{1/3}\,,
\end{equation}
where $I_{\rm c}$ is the companion star's rotational moment of inertia.
Following \citet{Zahn1977}, if the star rotates with an angular frequency
$\Delta \Omega_{\rm c}$ larger than the synchronous frequency $\Omega_0 \simeq
\Omega$, which we can approximate by the orbital frequency $\Omega$ due to the
much larger angular momentum of the orbit compared to the spin of the star, then
tidal forces will reduce $\Delta \Omega_{\rm c}$ to zero over the
\textit{synchronisation timescale},
\begin{equation}
  t_{\rm sync} = -\frac{\Delta \Omega_{\rm c}}{\dot{\Omega}_{\rm c}} = \frac{I_{\rm
      c}}{6 k_2 q^2 M_{\rm c} R_{\rm c}^2}\left(\frac{M_{\rm c} R_{\rm
      c}^2}{L_{\rm c}}\right)^{1/3}\left(\frac{A}{R_{\rm c}}\right)^6\,,
\end{equation}
where $L_{\rm c}$ is the star's luminosity. This expession assumes that the star
has a large convective envelope. Rearranging for $\dot{P}_{\rm orb}$, we find
$\dot{P}_{\rm orb} \sim 10^{-8} \,(k_2 / 10^{-3})\, (\Delta \Omega_{\rm c} /
\Omega)$. For the asynchronous rotation of the companion star to the redback
PSR~J1723$-$2837, \citet{vanStaden2016+SpottyRB} found $(\Delta \Omega_{\rm c} /
\Omega) \approx 3\times10^{-3}$. Adopting a similar asynchronicity here results
in an expected $\dot{P}_{\rm orb} \sim 3\times10^{-11}$, of similar
magnitude to that observed from J2039.

Unfortunately, we are unable to measure an asynchronicity of this
  magnitude with existing data. Assuming $\Delta \Omega_{\rm c}/\Omega =
  3\times10^{-3}$, the star will undergo one extra rotation with respect to the
  co-rotating orbital frame every $75$ days, therefore completing several cycles
  between each of our light curves. Since the numbers of such cycles in between
  our observations are unknown and unmeasurable, we cannot unambiguously measure
  the asynchronous rotation by comparing the spot longitudes in each of our
  light curves. Indeed, we cannot even be sure that the same spot is present on
  each of our observing epochs. The asynchronous rotation also cannot be
  measured from the orbital period variations seen in Section~\ref{s:timing}, as
  these effects are caused by variations in the internal structure, and can
  occur even in the case of a tidally-locked companion star.

Thus, a tempting (although highly speculative!) picture emerges, in line with
that proposed by \citet{vanStaden2016+SpottyRB}, where many of the variable
phenomena seen from J2039 are due to magnetic activity and asynchronous rotation
in the companion star. In this picture, the magnetic activity leads to large
star spots, explaining the asymmetry in the optical light curve, and to
quadrupole moment variations in the stellar envelope, explaining the short-term
orbital period variations. Asynchronous rotation of the spotted surface then
leads to the observed optical variability, and introduces a tidal force that is
responsible for the putative long-term increase in the orbital period. A large
stellar magnetic field would also be consistent with the synchrotron explanation
for the orbital gamma-ray modulation described in
Section~\ref{s:orbvar}. Investigating this picture in the future will require
several more years of timing measurements to confirm or refute the long-term
$\dot{P}_{\rm orb}$, and high-cadence (e.g. monthly) optical monitoring
to test for evidence for asynchronous rotation in the form of a
periodicity in the optical variability.

\subsection{Prospects for binary gamma-ray pulsar searches}
Over the course of the \textit{Fermi}-LAT mission, a number of candidate redback
systems, similar to J2039, have been discovered within unidentified LAT sources
\citep[e.g.,][]{Strader2014+J0523,Li2016+J0212,Linares2017+J0212,Salvetti2017+PSRCands,Halpern2017+J0838,Li2018+J0954,Swihart2020+J2333}. Our
detection of gamma-ray pulsations from J2039 shows that, with sufficiently
precise orbital constraints, gamma-ray pulsation searches are a viable method to
confirm their redback natures.

\begin{figure}
  \includegraphics[width=\columnwidth]{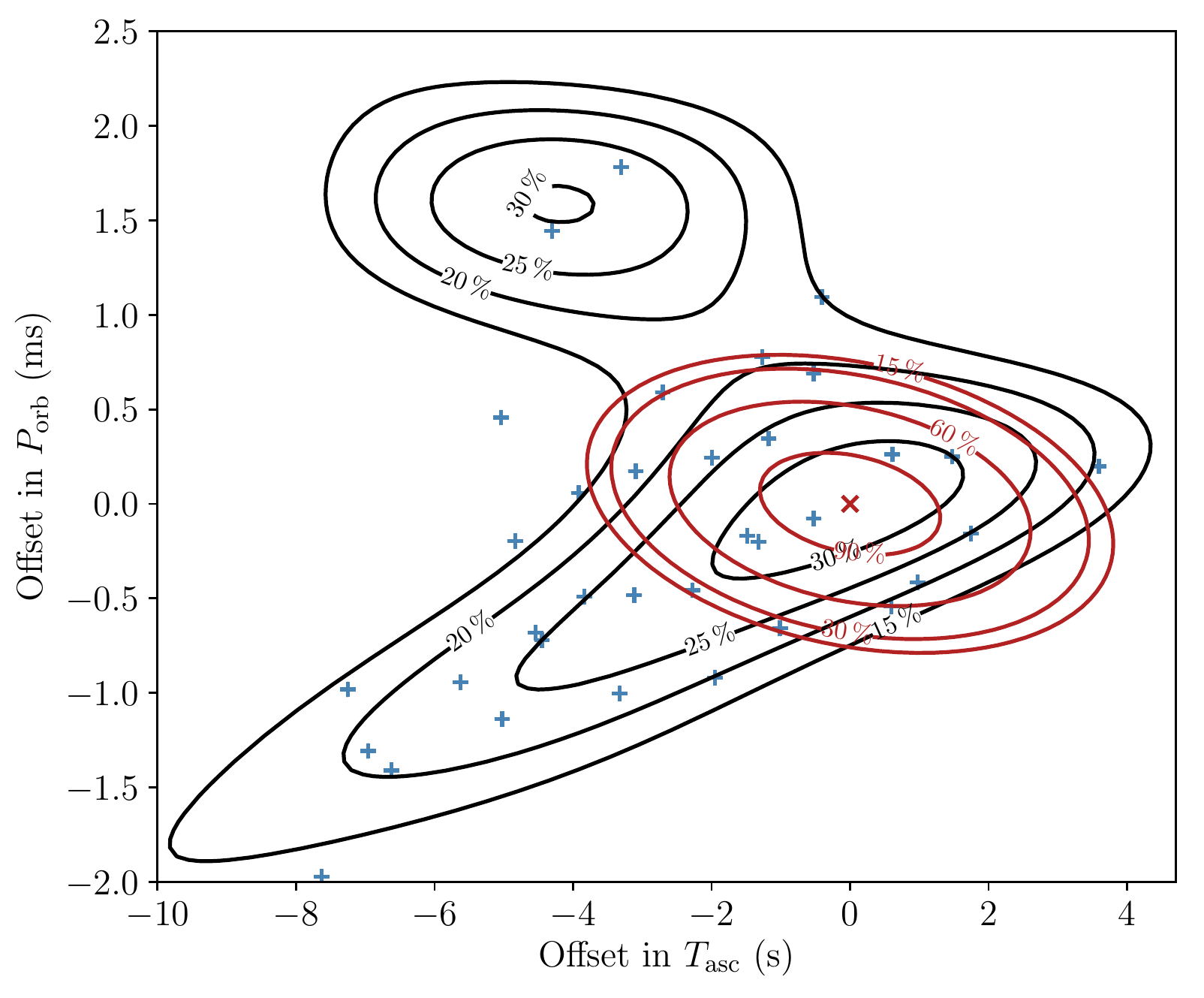}
  \caption{ \label{f:signalloss_PB_TASC} Fraction of maximum signal power
    recovered as a function of offsets in $P_{\rm orb}$ and $T_{\rm asc}$ from
    the timing solution in Table~\ref{t:params}. The origin is the point in
    parameter space giving the highest signal power for J2039 using a model for
    orbital motion with constant period. Red contour lines show the expected
    fraction of signal power recovered according to the metric approximation
    used to construct the search grid, which assumes that the signal has a
    constant orbital period. The effect of the orbital period variations is to
    reduce the maximum signal power, and to spread it over a
    larger region of the parameter space. The black contour lines
    show the actual recovered signal power as a function of $P_{\rm orb}$ and
    $T_{\rm asc}$. Blue crosses show parameter space locations at which a
    significant signal was detected in our search. Note that the $90\%$ ellipse
    was used for the grid generation in the search described in this work. A
    search grid designed to take into account the smearing effect of the orbital
    period variations could feasibly have been several times sparser in these
    parameters without missing the signal.}
\end{figure}

However, the orbital period variations common in redbacks, and present here for
J2039, will make them more difficult to detect. Indeed, J2039 is the first spider MSP exhibiting rapid orbital period
variations to have been discovered in a gamma-ray search. Due to the low photon
flux from a typical pulsar, multiple years of LAT gamma-ray data are required
for a discovery in a directed search. On such timescales the pulsar's ascending
node can shift back and forth by more than $15$\,s for some redback pulsars
\citep[see e.g.,][]{Pletsch2015+J2339,Deneva2016+J1048}. In a pulsation search,
we are forced to assume a constant orbital period. The fact that orbital period
variations are common in redbacks therefore has two major implications, which we
illustrate in Figure~\ref{f:signalloss_PB_TASC}.

Firstly, the recovered signal-to-noise ratio drops significantly. As shown in
Figure~\ref{f:signalloss_PB_TASC}, the maximum signal strength found in the
search for J2039 was $66\%$ smaller than the signal strength obtained after
accounting for the orbital period variations in our timing analysis. The reason
for this is visible from the left panel of Figure~\ref{f:OPV}, where it can be
seen that the signal becomes clearer as the offset in $T_{\rm asc}$ decreases,
while at epochs where the offset is largest ($\Delta T_{\rm asc} \sim \pm
10$\,s), the signal disappears entirely.
Despite this, J2039 was still easily detected above the
statistical noise level, but for fainter pulsars in future searches this could
reduce the signal strength below detectable levels.

Secondly, the signal is spread over a larger parameter volume compared to a
signal from a pulsar with a constant orbital period. This could actually be
beneficial to future searches: assuming the signal is strong enough to remain
detectable over small portions of the LAT data, the orbital period variations
may actually allow pulsations to be detected over a larger range in $P_{\rm
  orb}$ and $T_{\rm asc}$, as can be seen in Figure~\ref{f:signalloss_PB_TASC}. This could be exploited to reduce the computing cost
of future searches by using less dense grids over the orbital parameter
space. Another option could be to search the results for a clustering that
indicates a wider-than-expected spread of a signal. We intend to investigate
both options.

\section{Conclusions}
Using a directed search for gamma-ray pulsations running on the distributed
volunteer computing system \textit{Einstein@Home}, we have confirmed the
redback nature of the candidate binary system within 4FGL~J2039.5$-$5617. This
is the first redback pulsar to be discovered through its gamma-ray pulsations,
providing hope that a number of similar redback candidates identified in
\textit{Fermi}-LAT sources might be confirmed in this way in the near future, even though their orbital periods display large variability.

Optical observations of variations in the orbital light curve, and gamma-ray
timing observations of its changing orbital period, add another example to a
growing body of evidence that redback companions have activity on super-orbital timescales. A better understanding of variable phenomena in
redback companions is required both to ensure that the properties inferred from
optical light-curve modelling (e.g. inclination angles, pulsar and companion
mass estimates) are reliable, and to better
understand their evolution.

The origin of the light curve variability remains unclear, but requires
temperature variations of a few hundred K over a reasonably large fraction
of the visible surface of the star. We speculate on a few possible origins for
these temperature variations, including reprocessing of the pulsar's heating flux
in a variable intra-binary shock, variable convective flows on the stellar
surface, or magnetic star spot activity. The latter picture fits well with the
interpretation of orbital period variations being caused by quadrupole moment
variations driven by magnetic activity in the companion star.

To quantify the orbital period variations, we have developed a new gamma-ray
pulsation timing method that treats the orbital phase as a stochastically
varying function, and provides statistical estimates of the amplitude and
characteristic timescale of the variability. We find that the magnitude of the
orbital period variations requires only a small fractional change (a few
percent) in the stellar quadrupole moment, suggesting that this is indeed a
plausible scenario. However, due to the sparsity of optical observations, we are
so far unable to probe correlations between the optical light curve variability
and changes in the orbital period. Based on these phenomena, we are pursuing
long-term monitoring of redback companions to reveal whether or not optical
variability is correlated with quadrupole moment variations. Future
  light curves, ideally obtained with a denser (e.g. monthly) observing cadence,
  may also allow us to track variable light curve components as they evolve over
  time, providing a probe for possible asynchronous rotation.

We modelled the optical light curves, using the new timing measurement of the
projected semi-major axis of the pulsar's orbit to constrain the binary mass
function. Although our modelling is complicated and likely biased by the
unexplained variability and light-curve asymmetry, the gamma-ray data
significantly rule out any substantial eclipse and set a maximum inclination of
$i \lesssim 78\degr$, and we find that an inclination of $i\sim75\degr$ provides
a consistent fit to all light curves with an inclination as low as
$i\sim60\degr$ being consistent with one single-epoch light curve. This implies
a fairly low pulsar mass $1.1 M_{\odot} < M_{\rm psr} < $ 1.6 $M_{\odot}$, and
companion mass $0.15 M_{\odot} < M_{\rm c} < 0.22 M_{\odot}$. Additional
  light curves may help to reduce the inclination and mass uncertainties by
  ``marginalising out'' the variability, although these will remain highly
  model-dependent and subject to systematic biases without an independent method
  to validate with. The joint radio and gamma-ray pulse profile modelling
  described in Paper 2 is one such method, and broadly agrees with our results
  here.

We also find that an orbitally modulated component to the gamma-ray flux is in
fact pulsed emission in phase with the magnetospheric gamma-ray pulses, rather
than being an additional unpulsed component. We speculate that this could be due to inverse Compton scattering or synchrotron radiation from the high-energy pulsar wind. This is the second redback from which such an effect has been detected, and this may prove to be a valuable probe of the pulsars' high-energy winds in future studies.

\label{s:conclusions}

\section*{Acknowledgements}

We are very grateful to the thousands of volunteers who donated computing time
to \textit{Einstein@Home}, and to those whose computers first detected
PSR~J2039$-$5617: J. Bencin of Cleveland, OH, USA; and
an anonymous volunteer whose username is ``Peter''.
	
C.J.C. would like to thank Steven G. Parsons for guidance and useful discussions
while observing with ULTRACAM. 

C.J.C., G.V., R.P.B., M.R.K., D.M.-S. and J.S., acknowledge support from the ERC
under the European Union's Horizon 2020 research and innovation programme (grant
agreement No. 715051; Spiders). V.S.D. and ULTRACAM acknowledge the support of
the STFC. Z.W. is supported by the NASA postdoctoral
program. \textit{Einstein@Home} is supported by NSF grants 1104902 and 1816904.

The \textit{Fermi} LAT Collaboration acknowledges generous ongoing support
from a number of agencies and institutes that have supported both the
development and the operation of the LAT as well as scientific data analysis.
These include the National Aeronautics and Space Administration and the
Department of Energy in the United States, the Commissariat \`a l'Energie Atomique
and the Centre National de la Recherche Scientifique/Institut National de Physique
Nucl\'eaire et de Physique des Particules in France, the Agenzia Spaziale Italiana
and the Istituto Nazionale di Fisica Nucleare in Italy, the Ministry of Education,
Culture, Sports, Science and Technology (MEXT), High Energy Accelerator Research
Organization (KEK) and Japan Aerospace Exploration Agency (JAXA) in Japan, and
the K.~A.~Wallenberg Foundation, the Swedish Research Council and the
Swedish National Space Board in Sweden.

Additional support for science analysis during the operations phase is
gratefully acknowledged from the Istituto Nazionale di Astrofisica in Italy and
the Centre National d'\'Etudes Spatiales in France. This work performed in part under DOE Contract DE-AC02-76SF00515

Based on observations collected at the European Southern Observatory under ESO
programme 0101.D-0925(B). Based on data obtained from the ESO Science Archive Facility under request number 316850.

This paper uses data taken from the XMM-Newton Science Archive (Observation ID:
0720750301) and produced using the Pipeline Processing System. This work has
made use of data from the European Space Agency (ESA) mission {\it Gaia}
(\url{https://www.cosmos.esa.int/gaia}), processed by the {\it Gaia} Data
Processing and Analysis Consortium (DPAC,
\url{https://www.cosmos.esa.int/web/gaia/dpac/consortium}). Funding for the DPAC
has been provided by national institutions, in particular the institutions
participating in the {\it Gaia} Multilateral Agreement. This research was made
possible through the use of the AAVSO Photometric All-Sky Survey (APASS), funded
by the Robert Martin Ayers Sciences Fund and NSF AST-1412587.

\section{Data Availability}
The \textit{Fermi}-LAT data are available from the Fermi Science Support Center (\url{http://fermi.gsfc.nasa.gov/ssc}). The \textit{XMM-Newton} data are available from the XMM-Newton Science Archive (\url{http://nxsa.esac.esa.int}).
The ULTRACAM data are available in Zenodo (\url{https://doi.org/10.5281/zenodo.3964204}).
The GROND data are available from the ESO Science Archive Facility (\url{http://archive.eso.org/}).

\bibliographystyle{mnras}
\bibliography{ms} 

\bsp	
\label{lastpage}
\end{document}